\documentclass[11pt]{article}

\usepackage{float}
\usepackage{amsmath,amssymb,graphicx}
\usepackage{authblk}
\usepackage{cite}

\usepackage{setspace}
\usepackage{tocloft}

\usepackage{bm}
\usepackage{multirow}
\usepackage{color}

\usepackage{comment}

\begin{document} 

\title{Head-wearable Holographic Head-mounted Display with 6 Degrees of Freedom}

\author[1]{Taichi Sakakihara}
\author[1]{Teppei Jodo}
\author[1]{Seok Kang}
\author[1]{Yuji Sakamoto \thanks{Email: hologram.researcher.yuji@gmail.com}}

\affil[1]{Hokkaido university, Graduate school of information science and technology, N14 W9, Kita-ku, Sapporo, JAPAN}
\date{}

\renewcommand{\cftdotsep}{\cftnodots}
\cftpagenumbersoff{figure}
\cftpagenumbersoff{table} 

\maketitle

\begin{abstract}
A head-mounted display (HMD) using holography technology (holo-HMD) is expected to be the next generation of HMDs capable of reducing three-dimensional sickness.
In HMDs, it is important to generate images that respond to head movement in real time.
However, in holo-HMDs, generation of  hologram data in real time is difficult due to the large computational resources required.
This paper proposes a fast calculation algorithm for generating hologram data for holo-HMDs, which requires low computational power. A holo-HMD supporting six degrees of freedom was also developed using this algorithm and it was confirmed that it obtained reconstructed images with six degrees of freedom in real time (30 fps or more).

\end{abstract}


\section{Introduction}
\label{sect:intro} 


Head-mounted displays (HMDs) are commonly used as three-dimensional (3D) image-display devices for augmented reality (AR) and virtual reality (VR). 
An HMD for VR generates parallax images called stereogram by displaying parallax images in the left and right displays, which enables a viewer to see a 3D image.
In the stereogram, however, the perceptual depth of a viewer is different from the actual focal depth of the image, because an HMD displays images at the fixed depth determined with the optical system.
This mismatch in the two depths causes vergence accommodation conflict (VAC), which is one of the causes of eye fatigue and 3D sickness when using an HMD for long periods \cite{3Dsickness}.
Even with AR systems using monocular HMDs that do not display 3D images, frequent eye refocusing is required when the depth of the object in the real world differs from that of the optical image, which also causes eye fatigue.

Holography is considered an ideal 3D imaging technology because it can display images at any depth \cite{gabor1948, waters1966holographic}.
Electro-holography is an electronic implementation of the principles of holography, enabling video, electronic communication, and virtual scene display.
An HMD utilizing electro-holography is called a holographic HMD (holo-HMD), which is expected to reduce VAC and eye fatigue caused by focus adjustment of the eyes.
There have been many studies on holo-HMDs \cite{chang2020toward}.
Most involved benchtop-system prototypes for research into optical components or for principal purposes\cite{maimone2017holographic, jeong2019holographically, duan2020full, padmanaban2019holographic, chang2019holographic, chen2025penta}, but there have been investigated holo-HMDs that can actually be worn on the head \cite{takemori19973,murakami2017study,yoneyama2018}.


For VR and AR applications, to achieve a high sense of immersion with an HMD, it is important that the image changes in accordance with the user's head movement. There are two types of degrees of freedom in the HMD movement: three degrees of freedom, which corresponds to rotational movements of the head, and six degrees of freedom, which accounts for full motion including both rotation and translation. To achieve this, a holo-HMD must sense head movement and generate hologram data in real time. The hologram data are calculated using a computer using a computer-generated hologram (CGH) algorithm and displayed on spatial light modulators (SLMs).

However, it is difficult to generate hologram data in a short time, because it requires a huge amount of calculation. The computers attached to HMDs also generally have low computing power. Generating holograms that produce realistic reconstructed images requires even more computation time. It is thus difficult to generate hologram data in real time for holo-HMDs with three degrees of freedom or six degrees of freedom.

The final aim of our line of research is to explore the feasibility of practical holo-HMDs.
Their most critical challenge is achieving real-time hologram data generation for six degrees of freedom.
To address this, we propose a fast calculation algorithm for generating hologram data for holo-HMDs and that requires low computational power. We confirmed that our algorithm can be used to enable a holo-HMD to produce six degrees of freedom reconstructed images in real time.

\section{Related work}


There are various methods for calculating hologram data in CGH. We explain the point-light source method.

\subsection{Point-light method}

\begin{figure}[ht]
  \centering
    \includegraphics[clip,width=8cm]{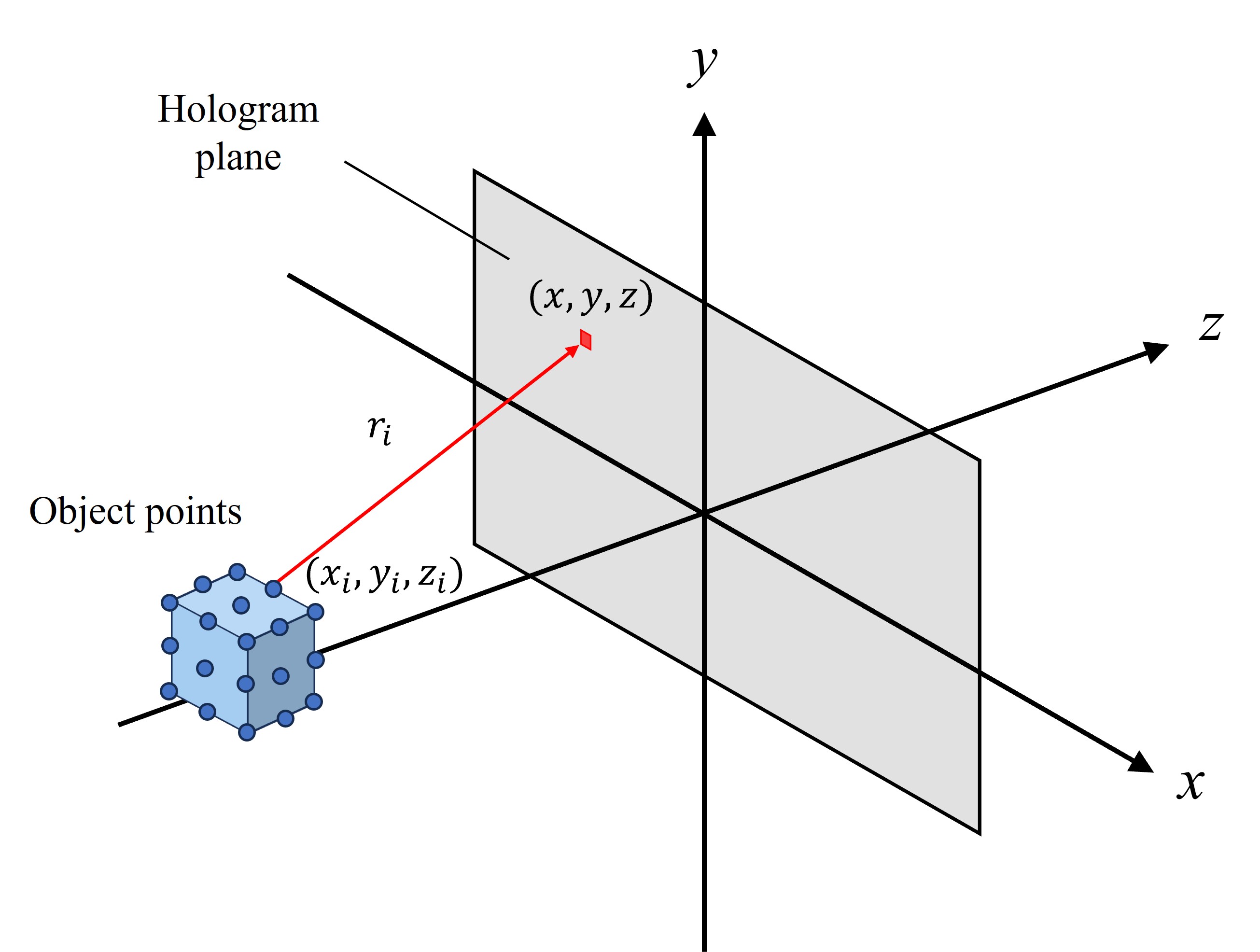}
    \caption{point-light method. An object is represented as a collection of point light sources, and object light on hologram plane is obtained as synthesis of spherical waves from each point light source.}
    \label{fig:point_base}
\end{figure}

We used the point-light method \cite{ogihara2015fast} to calculate light waves propagating from an object. This method represents a virtual object as a group of point light sources, and calculates the light-wave propagation from each point light source to the hologram. Finally, these contributions are summed to calculate the complex amplitude distribution of the object light.

Consider the object light wave distribution to a hologram of a virtual object defined by point light sources. When the coordinates of an arbitrary point light source are $(x_i,y_i,z_i)$ and an arbitrary pixel on the hologram is $(x,y,0)$, as shown in Fig. \ref{fig:point_base}, the object light wave $u_i$ propagating from the point light source to the hologram is expressed as
\begin{eqnarray}
  \label{eq:u_i}
  u_i(x, y) &= \frac{a_i}{r_i}\exp\{-j(kr_i+\phi_i)\}\\
  \label{eq:k}
  k &= \frac{2\pi}{\lambda}\\
  \label{eq:r_i}
  r_i &= \sqrt{(x-x_i)^2 + (y-y_i)^2 + z_i^2},
\end{eqnarray}
where $a_i$ is the amplitude of the point source, $r_i$ is the distance from the point light source to the pixel $(x,y,0)$ on the hologram, $k$ is the wavenumber, and $\phi_i$ is the initial phase of the point source.
When the total number of point sources of the virtual object is $N$, the light wave distribution $O(x,y)$ of the object light on the hologram is expressed as the sum of $u_i$ from each point source, as in the following equation.
\begin{equation}
    \label{eq:O(x,y)}
    O(x, y) = \sum^N_{i=1}u_i(x, y)
\end{equation}

This equation shows that the computational complexity of the point-light method is proportional to $N$ and the number of pixels in the hologram $P$.
As a result, the amount of calculation becomes extremely large, and a computer with high computational power is required.

The hologram data $I(x,y)$ are obtained by interfering the object light $O(x,y)$ from the reference light $R(x,y)$, as shown in
\begin{equation}
    \label{eq:I(x,y)}
    I(x, y) = \left| O(x,y) + R(x,y)\right|^2.
\end{equation}

\subsection{Fast calculation method}
\label{sec:fast}

Various methods have been proposed to speed up calculations with the point-source method. One important method is the table-look-up method \cite{lucente1993interactive, kim2008effective}, with which the object light from a single light source (Eq. 1) is calculated in advance then used to calculate Eq. 2, which is the superposition of the pre-calculated object light.
These methods achieved extremely fast calculations, but real-time calculations were not possible with the CPU of an ordinary PC.

By using general-purpose computing on graphics processing Units (GPGPUs) \cite{lee2014high, sakai2022autotuning}, fast calculations can be executed by taking advantage of the massive parallelism of the computing cores. This makes it possible to conduct real-time calculations even with a small number of point-light sources.
Currently, GPUs are commonly employed for the point source method, and the proposed algorithm was implemented on a GPU in this study.

However, calculating realistic rendering, which can represent hidden surface removal and object textures (shading) \cite{nishi2025rendering, watanabe2021hidden}, requires an even greater amount of calculation, making real-time calculations impossible. Methods for fast realistic rendering have been proposed using the ray tracing cores built into modern GPGPUs \cite{blinder2021photorealistic, watanabe2024realistic}, but real-time calculations have not yet been achieved. CGH calculations using deep neural networks (DNNs) have been proposed, making it possible to achieve extremely fast calculations \cite{DNN1, DNN2}, but there are problems with generating hologram data that correspond to changes in viewpoint.
Recently, real-time CGH generation that supports free-viewpoint rendering has been demonstrated by DNNs \cite{chen2025view,zhan2025complex}. However, these studies still exhibit inherent limitations, including phase discontinuities, insufficient physical consistency, scalability constraints, and anticipated difficulties in learning from real-world data. As a result, existing real-time free-viewpoint CGH methods cannot be applied to computations performed on a user-side PC.

Ray-tracing methods assume light rays emitted from the hologram plane and are capable of generating realistic images with effects such as hidden-surface removal and specular reflections \cite{watanabe2024realistic, ono2025fast}.
However, these methods require substantial computational power; on user-side systems with limited computational resources, the computation can take from several minutes to several hours. As a result, ray-tracing-based approaches are not suitable for achieving real-time free-viewpoint performance.
Since high-performance computing resources are available on the broadcasting-station side, the use of ray-tracing methods is feasible in this context.
\color{black}

\subsection{Broadcast Systems}
\label{sec:HVS}

One application of electro-holography is a 3D broadcast television system (Holo-TV), in which a broadcasting station generates and broadcasts wavefront data of a scene then viewed by multiple users on holo-HMDs or other electro-holographic display devices. When the broadcaster determines the viewpoint, generates hologram data, then transmits the hologram data, all users will view a 3D image from the same viewpoint, which the users has no degrees of freedom, 0DoF, such as in 3D movies.

\begin{figure}[ht]
  \centering
    \includegraphics[clip,width=12cm]{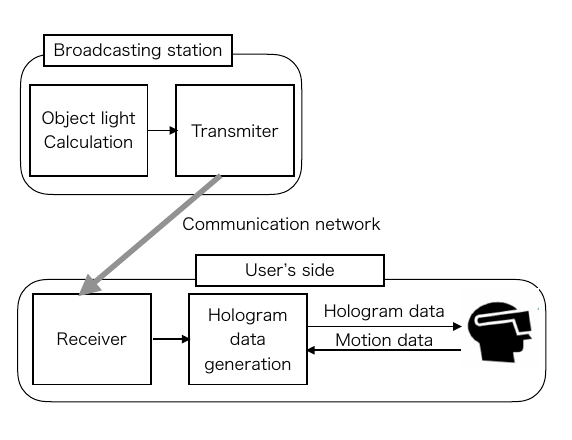}
    \caption{Schematic diagram of holometric video streaming (HVS) for broadcasting.}
    \label{fig:HVS}
\end{figure}
\color{black}

Holometric video streaming (HVS) has been proposed to achieve free viewpoints on the user side \cite{kon2024real, baba}. With this method, wavefront data, which requires a large amount of calculation power, are generated by the broadcasting station, and simple processing is executed on the user side with less calculation power (see Fig.\ref{fig:HVS}).

The object light $O(x,y)$ can be used to represent the wavefront data. Most of the computation time to generate a hologram data is spent on calculating the object light by using Eqs. \ref{eq:u_i} and \ref{eq:O(x,y)}, and the computational load from the object light to the hologram data by using Eq.\ref{eq:I(x,y)} is small. Therefore, even on low-computational-power terminals (holo-HMD), hologram data can be calculated quickly.\color{black}
In HVS, rather than calculating according to each user's head position and rotation, an object light with a larger area than the SLM is necessary.
From this object light, hologram data for a free viewpoint are generated according to each user's head movement.
This is similar to a user being able to view the outside world from various positions and angles through a large window.
This enables each user to simultaneously achieve a free viewpoint.

\section{Proposed holo-HMD}
\label{sec:Proposed method}

Our holo-HMD prototype has an enlarged field of view (FOV) and will be capable of full-color reconstruction for reconstructing images in real-time according to the user's head motion.

We introduce the three degrees of freedom of the user's head motion with our holo-HMD prototype. 
The user can rotate their head in three degrees of freedom: roll rotation when the head is tilted to the left or right, pitch rotation when the head is tilted up or down, and yaw rotation when the head is tilted horizontally.
It is possible to achieve six degrees of freedom with forward/backward, left/right, and up/down movements in addition to these three degrees of freedom.

\subsection{Optical system of prototype holo-HMD}
\label{sec:Optical system}

The half-width FOV $\theta_{max}$ of the reconstructed image displayed by an SLM is determined as
\begin{equation}
    \label{eq:theta_max}
    \theta_{max} = \sin^{-1}{\left( \frac{\lambda}{2p}\right)},
\end{equation}
where $p$ is the pixel pitch of the SLM.
Its FOV is very narrow for an HMD, so magnification using an optical system is required.

\begin{figure}[ht]
  \centering
    \includegraphics[clip,scale=0.4]{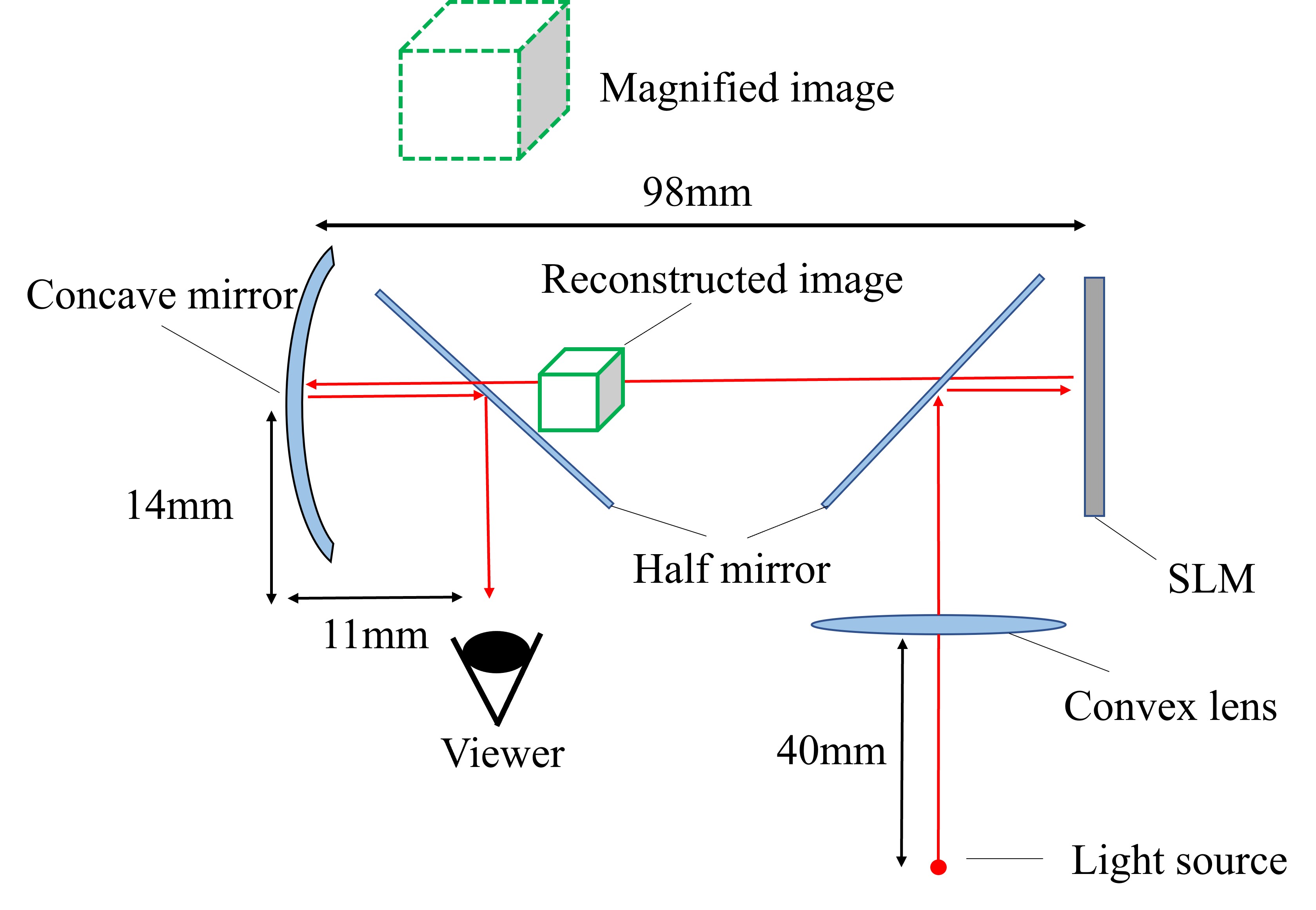}
    \caption{Schematic of optical system of our holo-HMD prototype. Laser light is collimated using convex lens and modulated using hologram displayed on SLM to form real image, and image is magnified using a concave mirror.}
    \label{fig:opt_sys}
\end{figure}

The optical system of our holo-HMD is shown in Fig. \ref{fig:opt_sys}. A laser beam from the light source is collimated using a convex lens and illuminated the SLM through a half-mirror as a reconstruction light. The diffracted light forms a reconstructed image, and the user observes a virtual image magnified using a concave mirror. Because a concave mirror is used to enlarge the reconstructed image, it is not affected by chromatic aberration for full-color reconstruction. 

A barrier is placed at the focal length of the concave mirror, which is the user's viewpoint. The zero-order light is focused onto a fixed point on and removed by the barrier. Higher-order diffraction images are removed at the same time. This optical system contributes to the miniaturization of our prototype because it does not require a long optical path as with the 4f optical system.

\subsection{Coordinate transformation for concave mirror}
\label{sec:concave mirror}

\begin{figure}[ht]
  \centering
    \includegraphics[clip,scale=0.45]{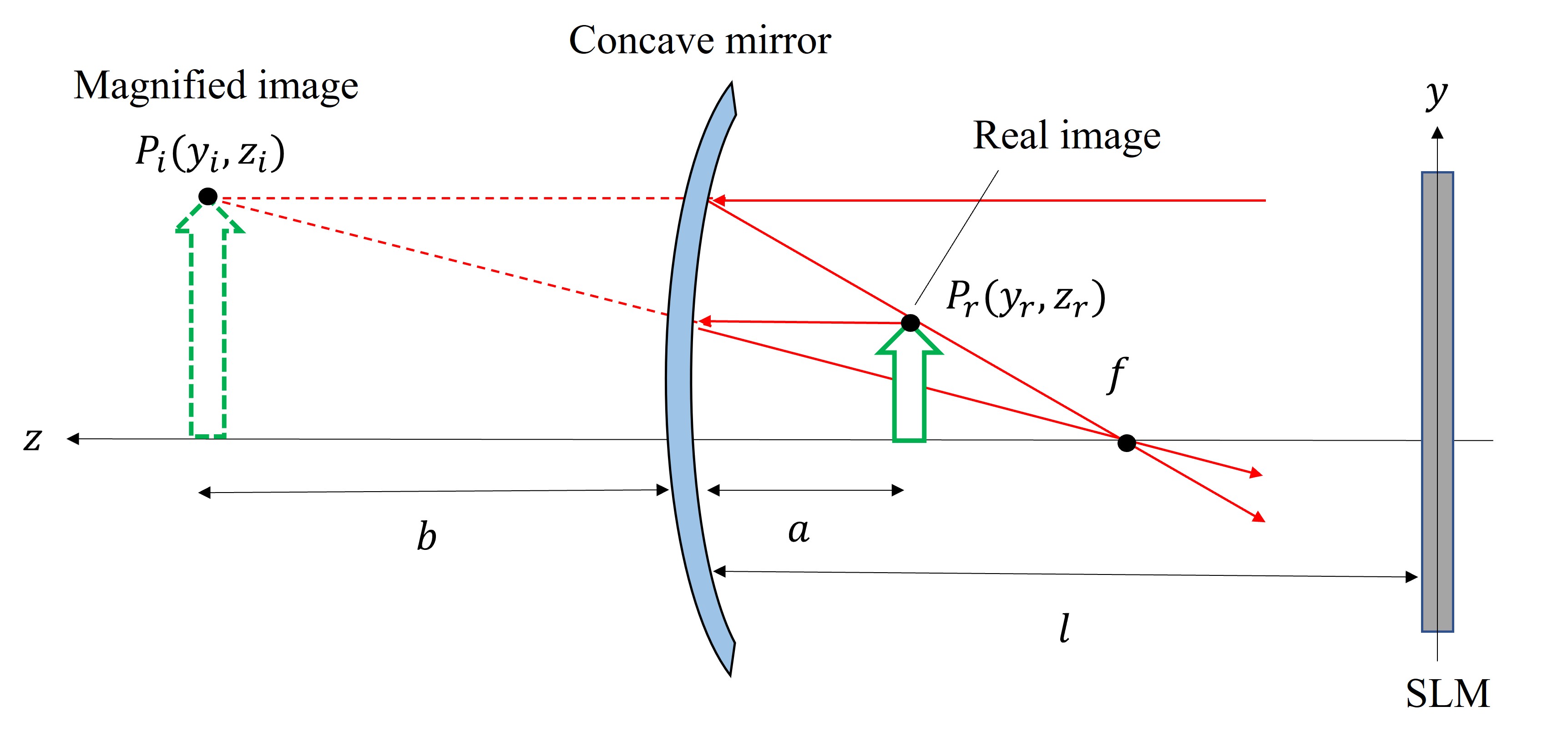}
    \caption{Magnification using concave mirror. Light wavefront modulated by the  SLM forms real image $P_r(y_r,z_r)$, which is magnified using concave mirror, such as magnifying glass, and magnified virtual image} $P_i(y_i,z_i)$ is viewed.
    \label{fig:convex}
\end{figure}

Since the concave mirror has the same magnification principle as a convex lens, a convex lens is used for this explanation (Fig. \ref{fig:convex}).

Because our holo-HMD prototype uses a concave mirror, the magnified image cannot be obtained at the correct position of the reconstructed image. The magnification ratio also increases with the coordinates closer to the focus of the concave mirror. Therefore, in the previous step of CGH calculation, the magnification effect of the concave mirror is taken into account to make the magnification rate constant, thus reconstructing the correct image.

Although the user observes the enlarged image, the object position specified during CGH calculation is usually the position of the reconstructed image before the enlargement. Therefore, it is necessary to obtain the coordinates of the reconstructed image before enlargement, which is necessary for CGH calculation, on the basis of the coordinates of the enlarged image that the user observes. Let $P_i(x_i,y_i,z_i)$ be the arbitrary coordinates of the reconstructed image after magnification and $P_r(x_r,y_r,z_r)$ be the arbitrary coordinates of the reconstructed image before magnification, the following formula is obtained from the lens formula for image magnification by using a concave mirror.

\begin{equation}
    \frac{1}{f} = \frac{1}{a} - \frac{1}{b}
\end{equation}
\begin{equation}
    \frac{1}{f} = \frac{1}{l-z_r} - \frac{1}{z_i-l},
\end{equation}
where $l$ is the distance between the concave mirror and SLM, and $f$ is the focal length of concave mirror. $z_r$ is expressed as
\begin{equation}\label{eq:z_r}
    z_r = \frac{l\{(z_i-l)+f\}-f(z_i-l)}{(z_i-l)+f}.
\end{equation}

From the lens formula, the magnification ratio $m$ is expressed as
\begin{equation}
    m = \frac{b}{a} = \frac{z_i-l}{l-z_r} = \frac{y_i}{y_r}.
\end{equation}

The following equation is used to organize $y_r$ and is expressed for the $x$-axis as well as the $y$-axis.
\begin{equation}\label{eq:y_r}
    y_r = y_i\frac{l-z_r}{z_i-l}
\end{equation}
\begin{equation}\label{eq:x_r}
    x_r = x_i\frac{l-z_r}{z_i-l}
\end{equation}

Finally, the coordinates $P_r(x_r, y_r, z_r)$ during CGH calculation are as follows.
\begin{equation}\label{eq:x_r,y_r,z_z}
    P_r(x_r,y_r,z_r) = \left(x_i\frac{l-z_r}{z_i-l},y_i\frac{l-z_r}{z_i-l},\frac{l\{(z_i-l)+f\}-f(z_i-l)}{(z_i-l)+f}\right)
\end{equation}


\subsection{Theoretical FOV}
\label{sec:FOV}

\begin{figure}[ht]
  \centering
    \includegraphics[clip,width=12cm]{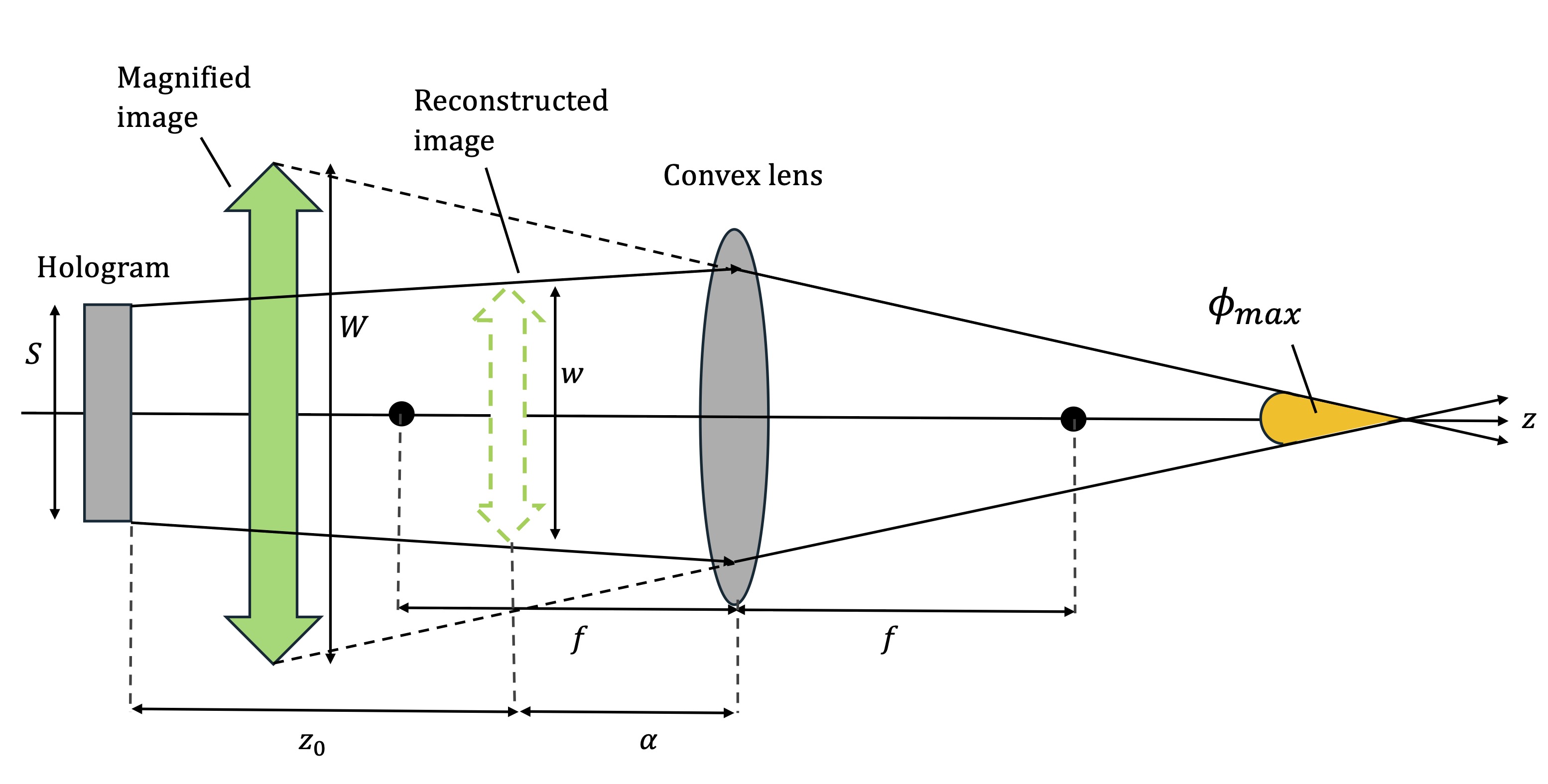}
    \caption{Schematic of optical system and theoretical maximum FOV of our holo-HMD prototype. An SLM generates light that forms a reconstructed image, which is magnified the a convex lens.}
    \label{fig:VOF-theory}
\end{figure}

As shown in Fig. \ref{fig:opt_sys}, light modulated using an SLM forms a reconstructed image between the SLM and concave mirror, which is then magnified by the mirror. The expansion of the FOV of an optical system with this configuration was discussed by Kikuchi et al. \cite{Kikuchi:25}. As shown in Fig. \ref{fig:VOF-theory}, by replacing the concave mirror in the holo-HMD with an optically equivalent convex lens, a theoretical FOV formula of the holo-HMD can be derived on the basis of Kikuchi's study. Let $f$ be the focal length of convex lens, $a$ be the distance between the lens and reconstructed image, $z_o$ be the distance between the lens and magnified image, $\lambda$ be the wavelength, $p$ be the pixel pitch of the SLM, and $S$ be the SLM size. The FOV of the holo-HMD can then be expressed as
\begin{equation}
    \phi_{max} =2\tan^{-1}\left(\frac{(z_0-f+a)w+(f-a)S},{2z_0f}\right),
\end{equation}
where 
\begin{equation}
    w =S+z_o \tan {\left(\sin^{-1}{\frac{p}{\lambda}}\right)}.\nonumber
\end{equation}
\color{black}

\subsection{six degrees of freedom fast-calculation algorithm}
\label{sec:algorithm}

Fast real-time calculation is required to achieve a six degrees of freedom display in accordance with the head motion. However, with the point-light method, the calculation time is proportional to the number of point light sources, making it difficult to calculate a CGH in real-time. Therefore, we introduce a fast-calculation algorithm to achieve six degrees of freedom at each user's display device and is, based on the HSV system discussed in Section \ref{sec:HVS}.

\begin{figure}[ht]
  \centering
    \includegraphics[clip,scale=0.4]{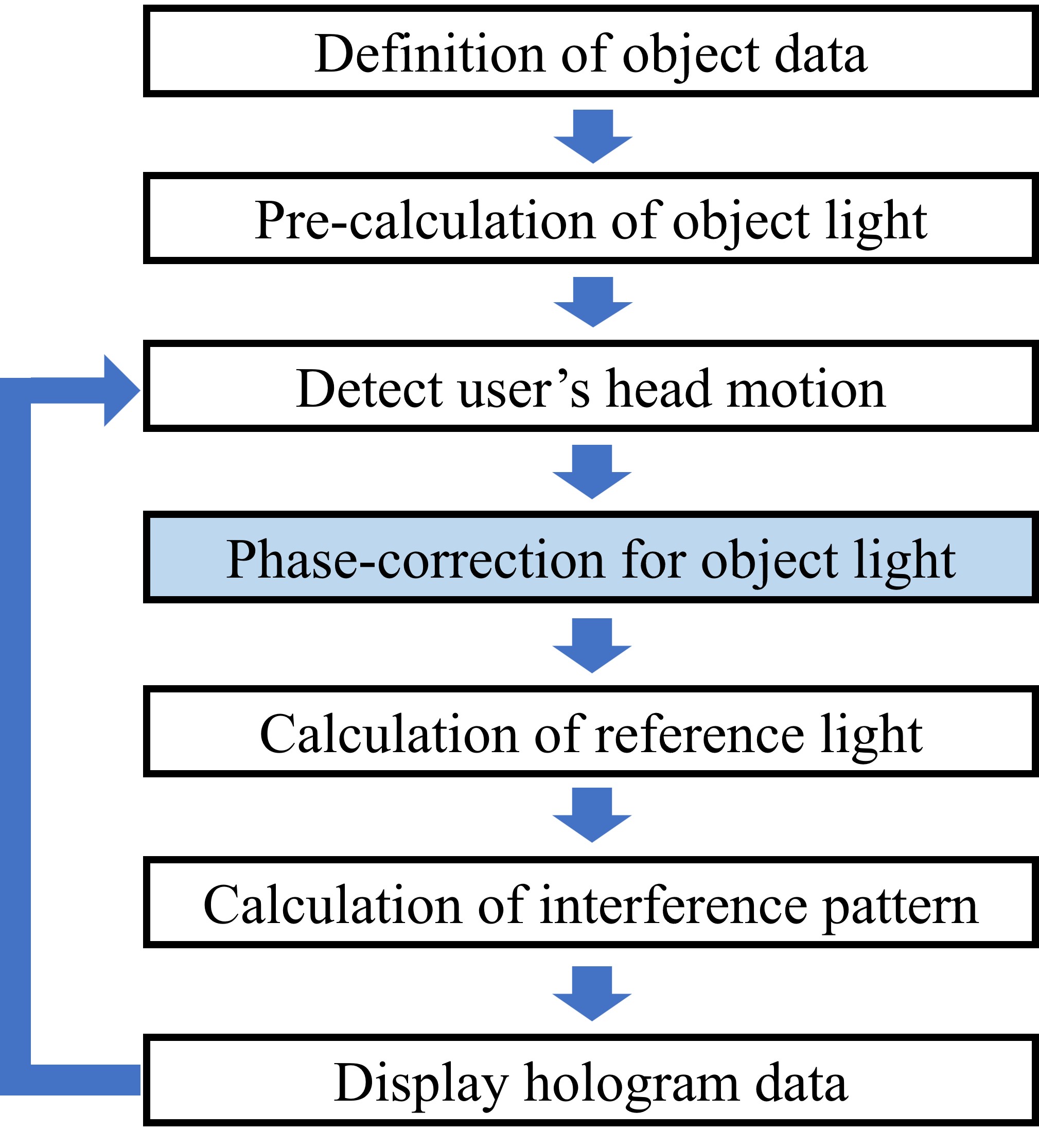}
    \caption{Fast-calculation algorithm. Interference patterns are obtained from pre-calculated object light by conducting phase-correction in accordance with user's head movement.}
    \label{fig:algorithm}
\end{figure}

We use phase-correction for our fast-calculation algorithm for real-time six degrees of freedom display (Fig. \ref{fig:algorithm}). The object light propagating from each point light source is first pre-calculated on the basis of the definition of virtual objects. This corresponds to the object light broadcast from the broadcaster in the HSV system. Next, the pre-calculated object light pixels to be referenced in each pixel of the hologram surface are determined according to the user's head motion. On the basis of the optical path difference $l$ between the pixels, the phase-correction described in the next section is then executed. Finally, interference patterns are calculated from the object light and reference light obtained after phase-correction, and a CGH is calculated.

By repeating the above calculations, the CGH can be calculated at high-speed in accordance with the user's head motion. With this method, once the object light pre-calculation is completed, the CGH is generated only by phase-correction, so there is no need to calculate the object light each time. The phase-correction does not depend on the number of light sources, so the calculation cost is very small. After the CGH is calculated, the image can be reconstructed in real-time by displaying it directly on the SLM via the graphics processing unit (GPU), instead of storing it in local storage.

\subsubsection{Object light phase-correction}
\label{sec:calc_order}
\begin{figure}[ht]
  \centering
    \includegraphics[clip,scale=0.38]{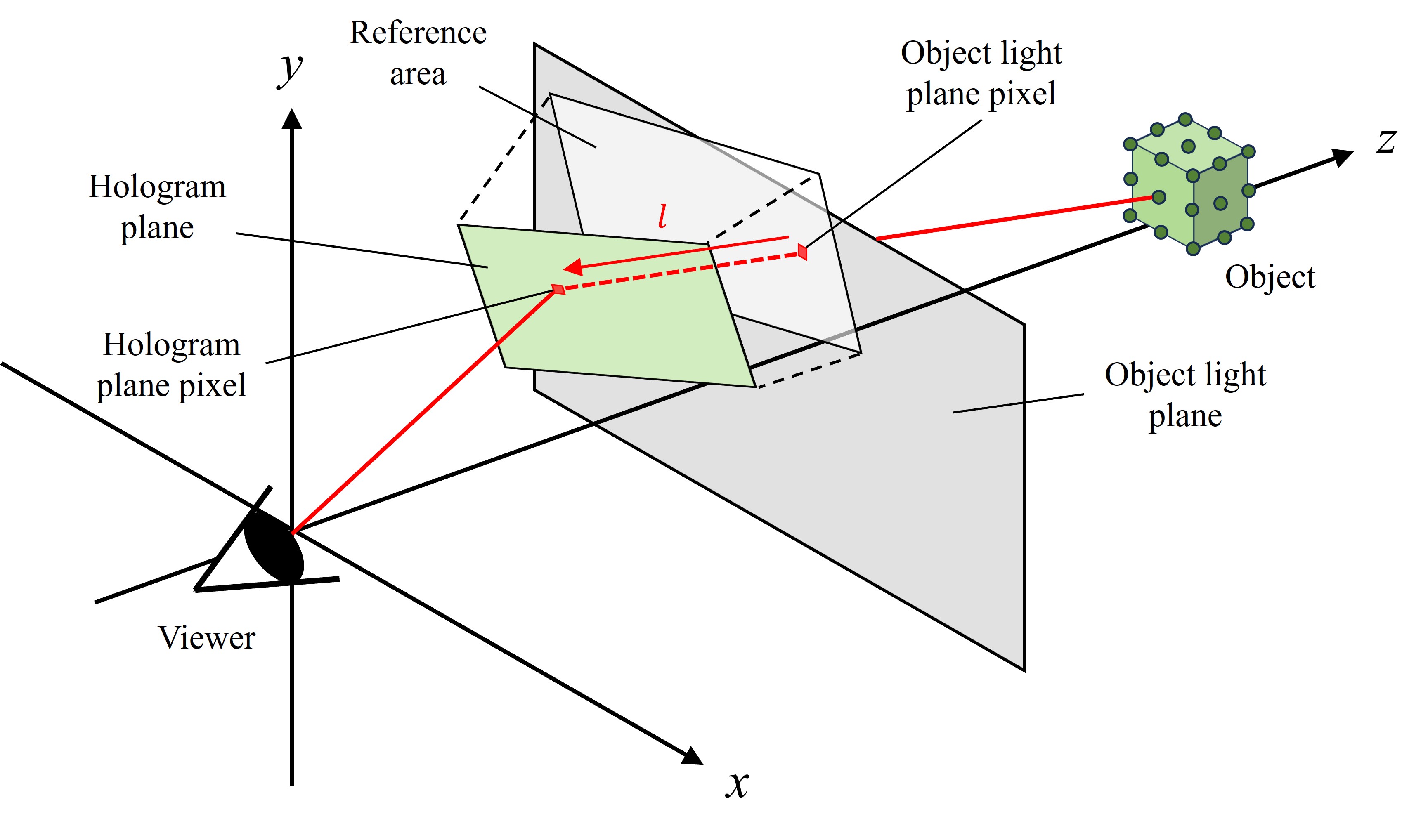}
    \caption{
    Propagation of light wavefront from pre-calculated object light to hologram plane is calculated as phase-correction.}
    \label{fig:phase_correction}
\end{figure}

\begin{figure}[ht]
  \centering
    \includegraphics[clip,scale=0.45]{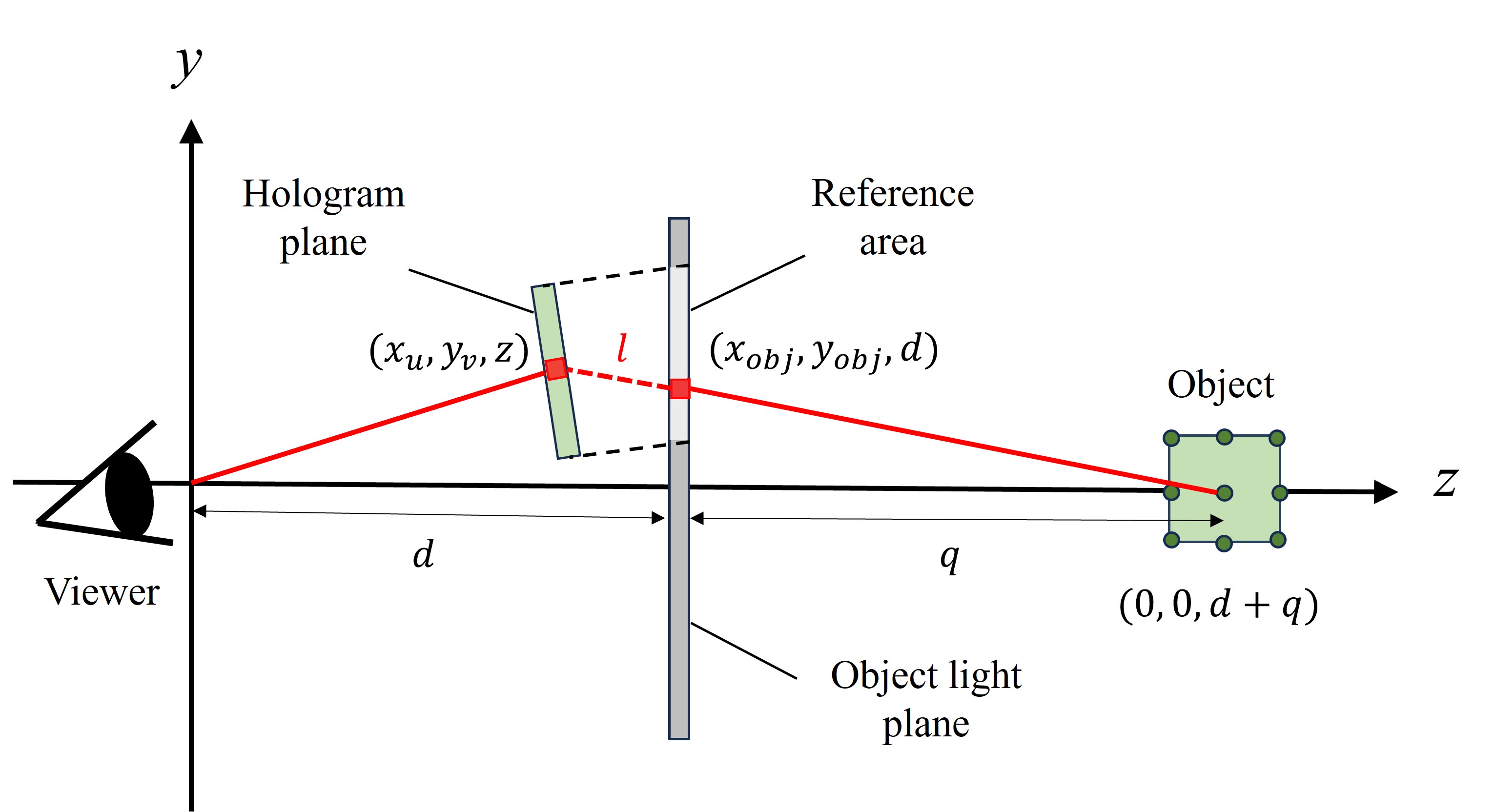}
    \caption{Phase-correction($x-z$ axis). Propagation of light wavefront from pre-calculated object light to hologram plane is calculated as phase-correction.}
    \label{fig:x_phase_correction}
\end{figure}

Schematic diagram of phase-correction is shown in Fig.s \ref{fig:phase_correction} and \ref{fig:x_phase_correction}. The object light to be pre-calculated is called the object light plane, and the distance from the object light plane to the virtual object defined during CGH calculation is $q$. The distance from the user's viewpoint to the object light plane $d$ is expressed with the following equation using the focal length $f$ and the distance between the SLM and concave mirror $L$ when there is no user head motion.
\begin{equation}
    d = f+L
\end{equation}
When the user's viewpoint changes in six degrees of freedom, the coordinates on the hologram plane to be obtained are determined with the rotation from the pre-calculated coordinates on the object light plane. If the rotation angle around the $z$-axis is $\theta_{roll}$, rotation angle around the $x$-axis is $\theta_{pitch}$, rotation angle around the $y$-axis is $\theta_{yaw}$, and translation vector is $\bm{t}$, then the coordinates $H(x_u, y_v, z)$ can be expressed as
\begin{equation}\label{eq:rot-mat}
    \begin{pmatrix}
       x_u \\
       y_v \\
       z
    \end{pmatrix}
     = R_z(\theta_{roll})R_y(\theta_{yaw})R_x(\theta_{pitch})
    \begin{pmatrix}
       x_0 \\
       y_0 \\
       z_0
    \end{pmatrix}
    +
    \begin{pmatrix}
       t_x \\
       t_y \\
       t_z
    \end{pmatrix}
\end{equation}
\begin{equation}\label{eq:(x_c,y_c,z_c)}
    (x_0,y_0,z_0) = \left(p\left(u-\frac{SLM_{width}}{2}\right),p\left(v-\frac{SLM_{height}}{2}\right),d\right),
\end{equation}
where $R_z(\theta_{roll})$ represents a rotation matrix of $\theta_{roll}$ around the $z$-axis, and similar rotation matrices in $R_y(\theta_{yaw})$ and $R_x(\theta_pitch)$. Coordinates $(x_0,y_0,z_0)$ are those before rotation on the object light plane and calculated on the basis of the SLM pixel pitch $p$ and the number of pixels $SLM_{width}$ and $SLM_{height}$. The coordinates $(x_obj, y_obj, z_obj)$ of the object light plane to be referenced are those passing through the two points $(x_u, y_v, z)$ and $(0, 0, d+q)$ and satisfying $z=d$, and are expressed with the following equation with parameter $s$.
\begin{equation}\label{eq:plane-object-light}
    (x_{obj},y_{obj},z_{obj}) = (tx_u,ty_u,d) 
\end{equation}
\begin{equation}
    s = \frac{(d+q)-d}{z-(d+q)}
\end{equation}

From Equations \ref{eq:rot-mat} and \ref{eq:plane-object-light}, the $l$ between the coordinates on the hologram to be obtained and those on the object light plane to be referenced can be expressed as
\begin{equation}
    l = \sqrt{x_u^2+y_v^2+(z-(d+q))^2}-\sqrt{x_{obj}^2+y_{obj}^2+(d-(d+q))^2}.
\end{equation}

When $l$ is sufficiently small, it can be regarded as the phase difference between two coordinates $(x_u, y_v, z)$ and $(x_{obj}, y_{obj}, z_{obj})$. Therefore, the object light wave distribution $O_h$ at $(x_u, y_v, z)$ on the hologram to be obtained can be expressed with the following equation using the pre-calculated light wave distribution $O_r$ on the object light plane and $k$.
\begin{equation}
    O_h = O_r\exp(-jkl)
\end{equation}

The above calculations are conducted for all pixels on the hologram to calculate the interference fringes. This phase-correction does not depend on the number of point light sources $N$ on the object but only on the number of pixels on the hologram $P$, so the calculation time is constant and fast even when $N$ is increased using the point-light method. The theoretical computational complexity of the point-light method is $O(PN)$. Even when $P$ is constant, the calculation time also increases $N$ increases. However the computational complexity of the phase-correction is $O(P)$ because it is determined only from $P$, indicating that the calculation time remains constant even if $N$ increases when $P$ is constant.


\subsection{Sampling theorem for phase-correction}
\label{sec:phase-correction}

We now find the maximum angle that can be approximated in phase-correction. For simplicity, we consider the horizontal yaw rotation. 

\begin{figure}[ht]
  \centering
    \includegraphics[clip,scale=0.45]{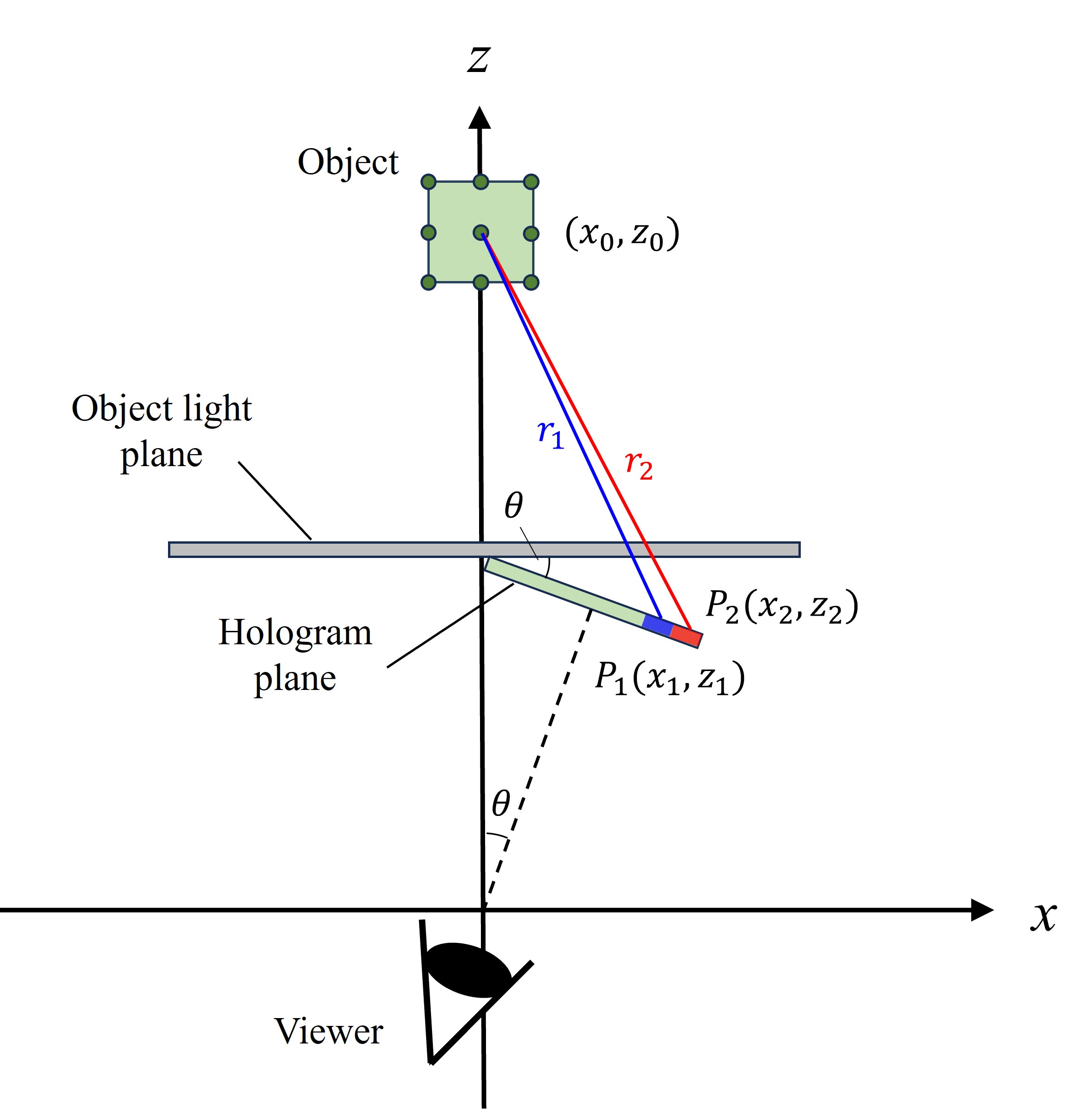}
    \caption{
Theoretical limits in phase-correction ($y-z$-axis).}
    \label{fig:y_phase_correction}
\end{figure}

A schematic diagram of the phase-correction viewed from the $y$-axis direction is shown in Fig. \ref{fig:y_phase_correction}. Let $r_2$ be the optical path length from the object center $(x_0,z_0)$ to hologram pixel $P_2(x_2,z_2)$, and $r_1$ be the optical path length to the adjacent hologram pixel $P_1(x_1,z_1)$.
\begin{equation}
    r_1 = \sqrt{(x_1-x_0)^2+(z_1-z_0)^2} \approx (z_1-z_0)+\frac{(x_1-x_0)^2}{2(z_1-z_0)}
\end{equation}
\begin{equation}
    r_2 = \sqrt{(x_2-x_0)^2+(z_2-z_0)^2} \approx (z_2-z_0)+\frac{(x_2-x_0)^2}{2(z_2-z_0)},
\end{equation}
where $r_1$ and $r_2$ are simplified using the Fresnel approximation. The coordinates of $P_1(x_1,z_1)$ and $P_2(x_2,z_2)$ are obtained from the following equations using the pixel $W$, rotation angle $\theta$, and pixel pitch $p$ of the hologram.
\begin{equation}
    P_1(x_1,z_1) = \left(\frac{(W-1)p}{2}\cos\theta,\frac{(W-1)p}{2}\sin\theta\right)
\end{equation}
\begin{equation}
    P_2(x_2,z_2) = \left(\frac{Wp}{2}\cos\theta,\frac{Wp}{2}\sin\theta\right)
\end{equation}

When calculating the approximation of holograms by phase-correction, it is necessary to consider the occurrence of aliasing due to resampling. Therefore, $\theta$ is determined under the condition that the optical path difference between adjacent hologram pixels $|r_1-r_2|$ satisfies the sampling theorem expressed as
\begin{equation}
    |r_1-r_2| < \frac{\lambda}{2}.
\end{equation}

\section{Implementation}

\subsection{Overview of holo-HMD prototype}
\begin{figure}[ht]
  \centering
    \includegraphics[clip,scale=0.5]{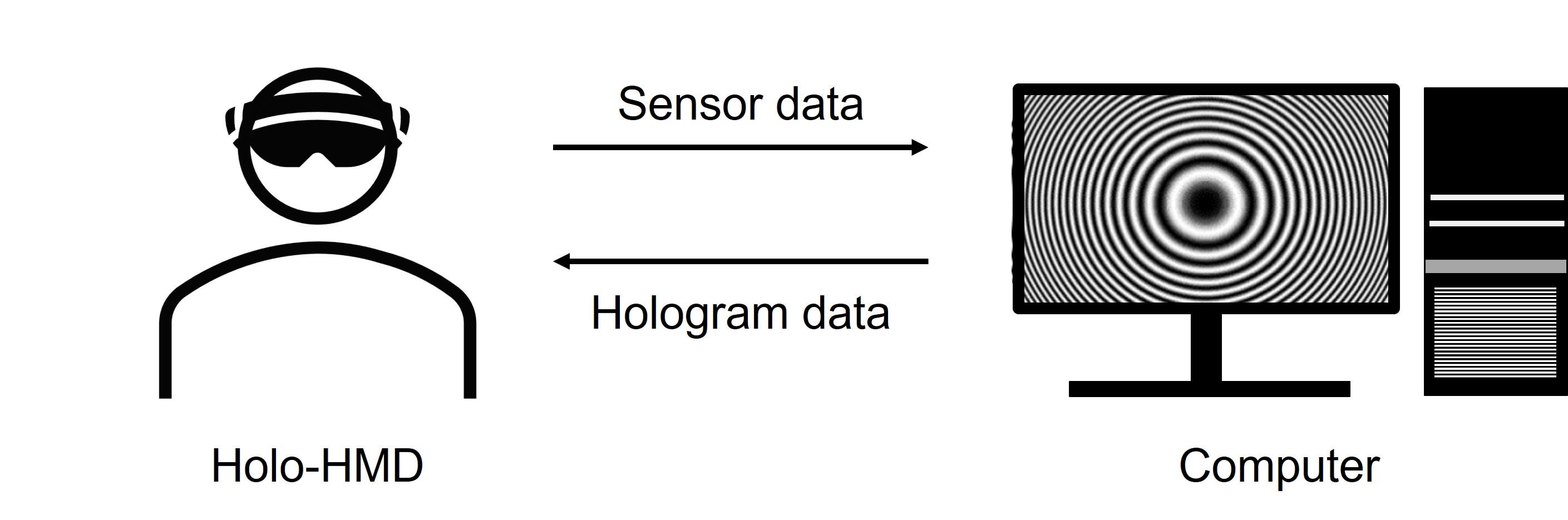}
    \caption{Proposed holo-HMD system.}
    \label{fig:Holo-HMD_system}
\end{figure}

\begin{figure}[ht]
    \begin{center}
        \begin{tabular}{cc} 
            \includegraphics[scale=0.22]{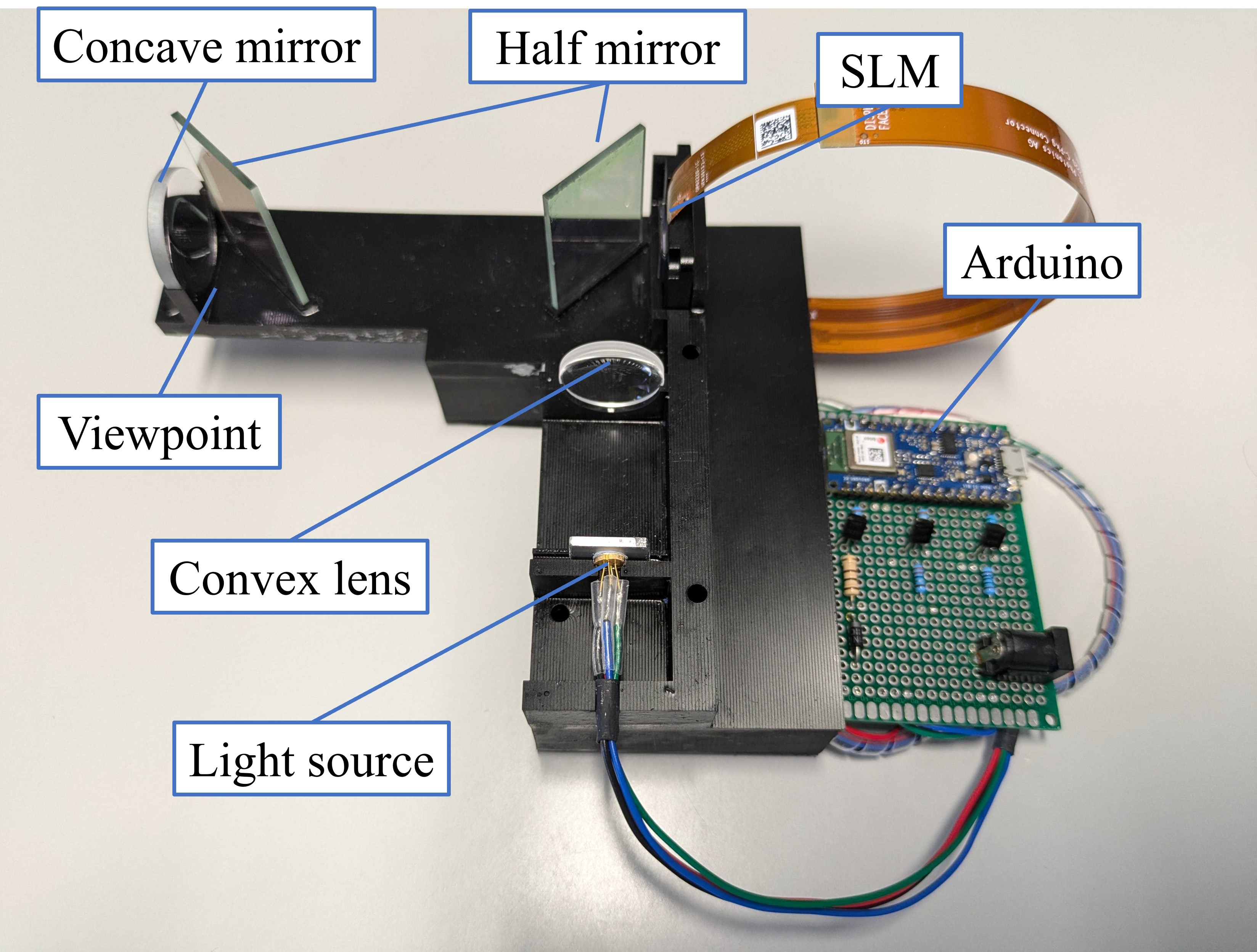} &
            \includegraphics[scale=0.22]{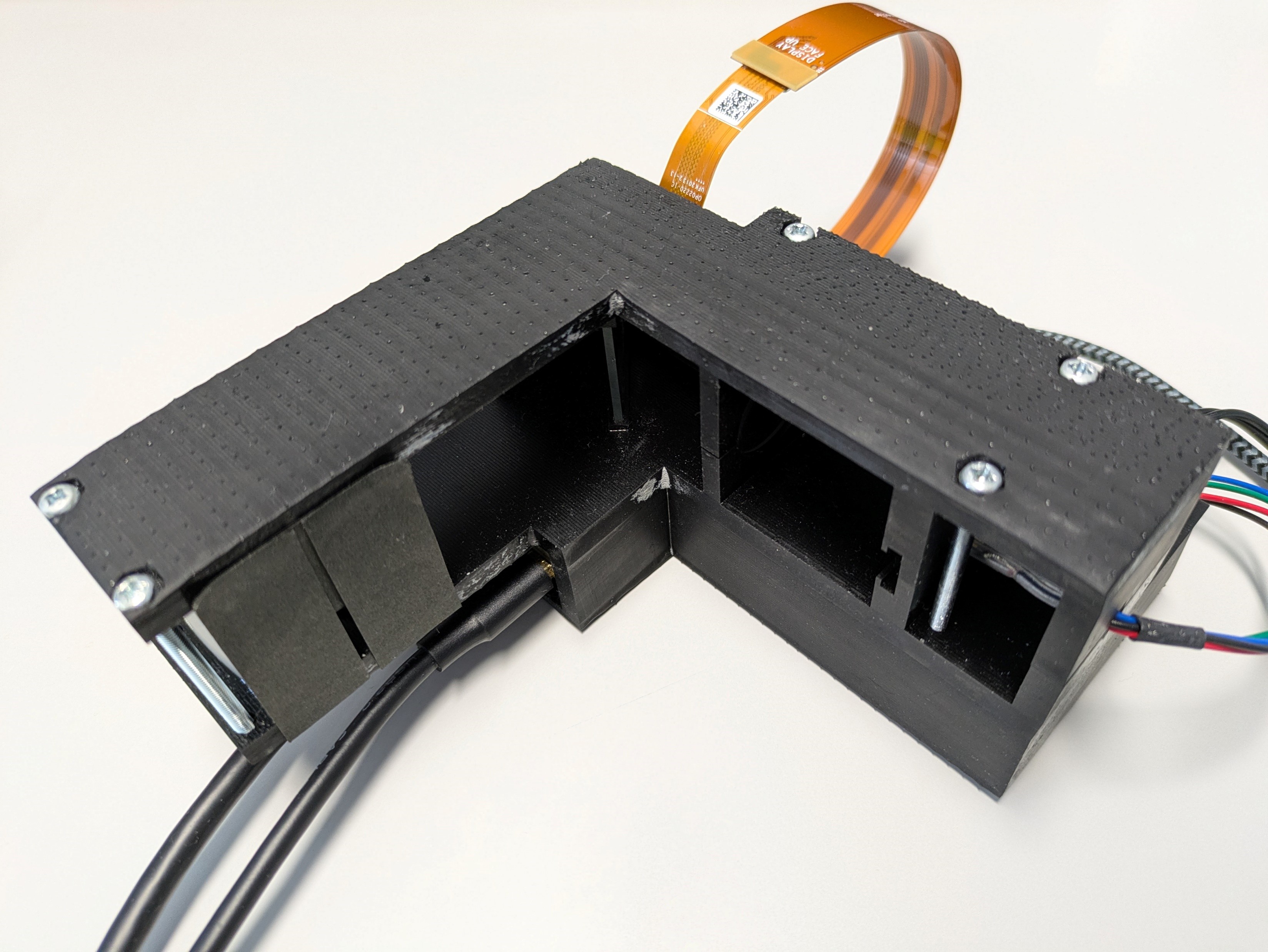} \\
            (a) & (b)
        \end{tabular}
    \end{center}
\caption{Photos of holo-HMD prototype: (a) interior, (b) exterior}
\label{fig:Holo-HMD}
\end{figure}

\begin{figure}[ht]
  \centering
    \includegraphics[clip,scale=0.3]{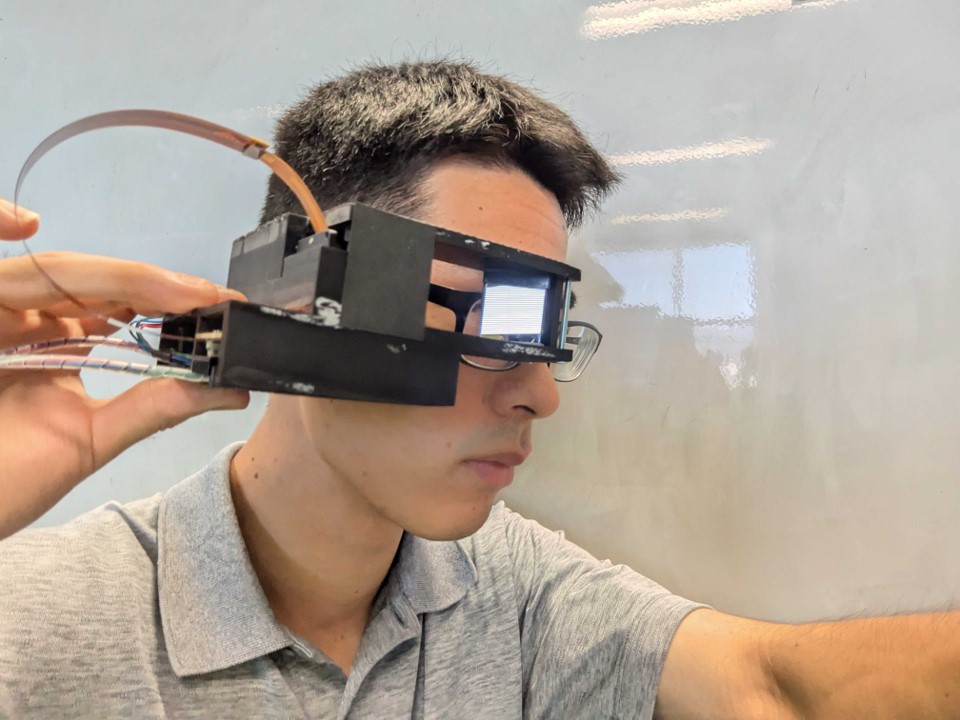}
    \caption{Our holo-HMD prototype in use.}
    \label{fig:HMD_usage}
\end{figure}

A schematic of our holo-HMD prototype is shown in Fig. \ref{fig:Holo-HMD_system}, and the photos of it are shown in Fig. \ref{fig:Holo-HMD}. Our holo-HMD prototype in use is shown in Fig. \ref{fig:HMD_usage}, and the device parameters are listed in Table \ref{tab:HMD_parameters}. 
The resolution of the SLM used is full high definition with a pixel pitch of $4.5 \mu m$. Our prototype is small, about the size of the palm of the hand, and weighs 287 g (monocular), within 600 g if used for a binocular system. This weight is equivalent to that of VR HMDs that have been commercialized, making it practical.

In our holo-HMD prototype, misalignment of optical components affects the observation of reconstructed images, so a high degree of precision is required. Therefore, we aimed to achieve high precision by using a stereolithography 3D printer.

Our holo-HMD prototype detects the user's head motion using the motion sensor built into the Arduino, sends the motion data to the computer, and executes real-time CGH calculation and displays it on the SLM on the computer. This enables real-time calculation and display of the CGH according to the user's viewpoint. If the holo-HMD system does not support six degrees of freedom, when the user moves their head while wearing the holo-HMD, the image also moves, causing it to behave differently from the real-object in the AR display. Therefore, it is desirable to switch the CGH in real-time according to the user's head motion, so that the image does not move even if the user moves their head, but appears to remain in place just like a real object.

\begin{table}[ht]
    \centering
    \caption{Device parameters of holo-HMD prototype.}
    \vspace{2ex}
    \label{tab:HMD_parameters}
    \begin{tabular}{llc}
        \hline 
        Size & $W \times D \times H$ & $125 \times 130 \times 40$ [mm] \\
        \hline 
        Weight & & $287$ [g] \\
        \hline 
        \multirow{2}{*}{Focal length} & Convex lens & $40$ [mm] \\
        & Concave mirror & $-25$ [mm] \\
        \hline
        \multirow{5}{*}{SLM} & Pixel pitch & $4.5 \times 4.5$ [{\textmu}m] \\
        & Resolution & $1920 \times 1080$ [pixels] \\
        & Display size & $8.64 \times 4.86$ [mm] \\
        & Frame rate & $60$ [Hz] \\
        & Color field rate & $360$ [Hz] \\
        \hline
        \multirow{3}{*}{Wavelength} & Red & $638$ [nm] \\
        & Green & $518$ [nm] \\
        & Blue & $448$ [nm] \\
        \hline
    \end{tabular}
\end{table}

\subsection{Full-color system}
\begin{figure}[ht]
    \begin{center}
        \begin{tabular}{cc} 
            \includegraphics[scale=0.3]{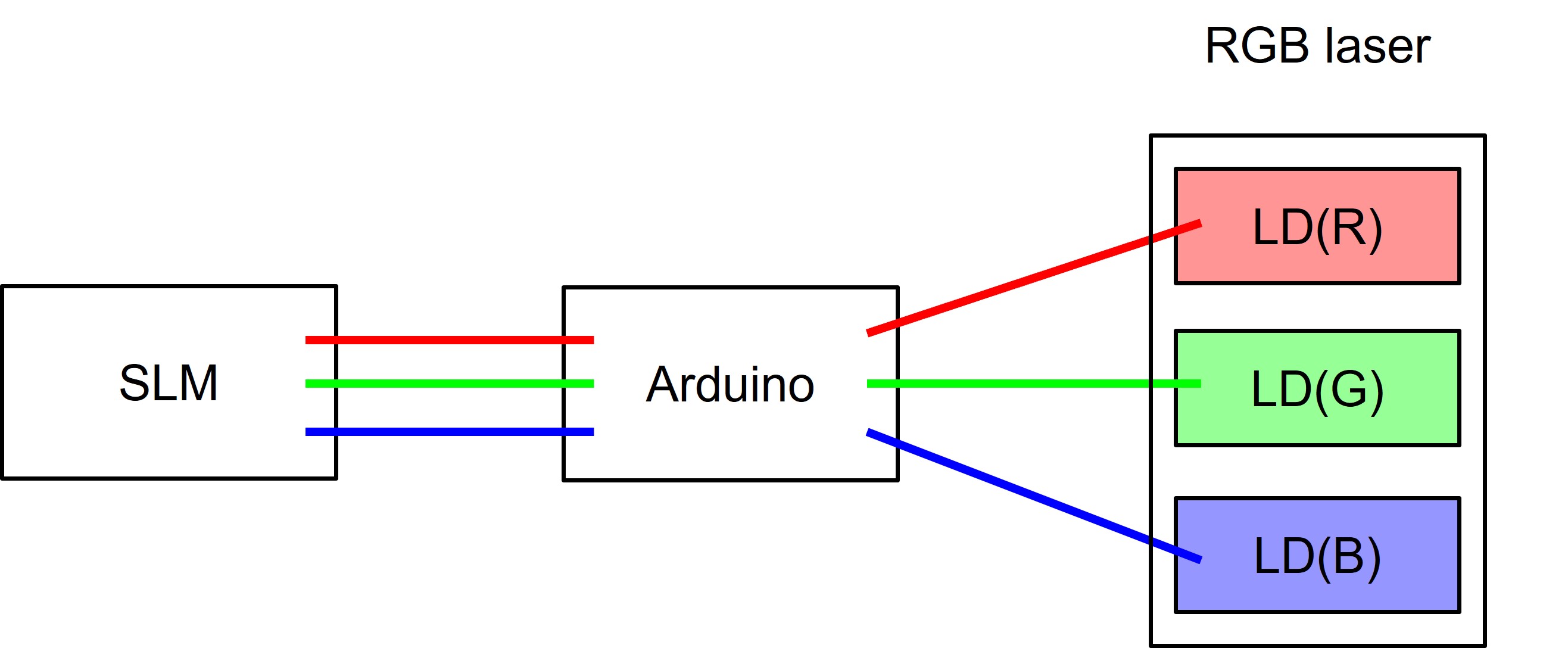} &
            \includegraphics[scale=0.3]{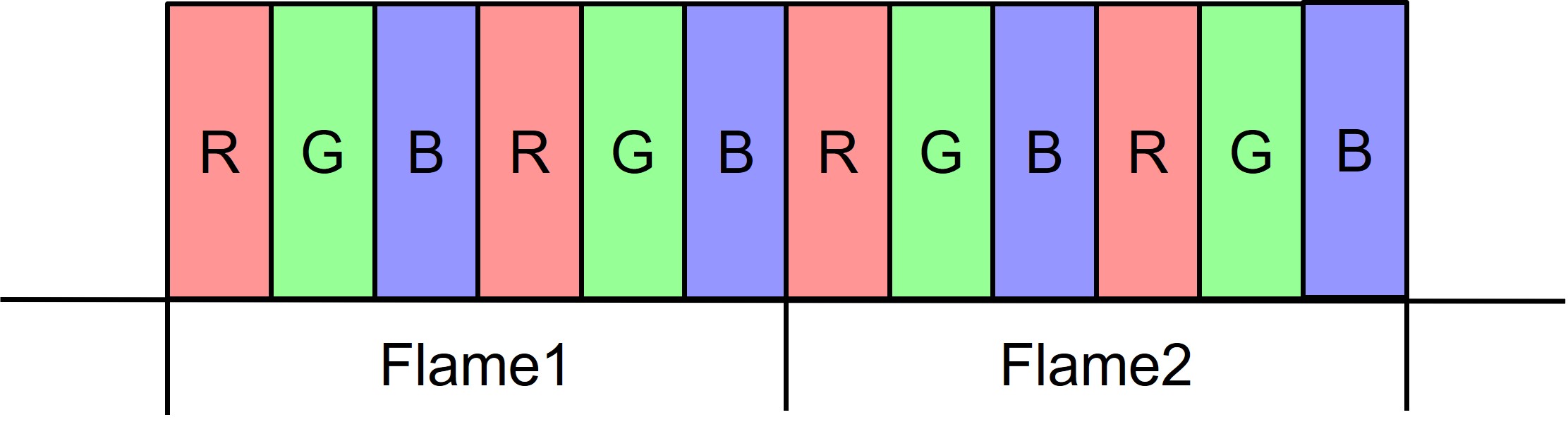} \\
            (a) & (b)
        \end{tabular}
    \end{center}
\caption{Full-color system: (a) full-color synchronization, (b) time-division multiplexing}
\label{fig:full_color}
\end{figure}

The full-color system of our holo-HMD prototype is shown in Fig. \ref{fig:full_color}a. The prototype uses a time-division multiplexing system to reconstruct full-color images, which enables full-color reconstruction using a single laser while keeping the holo-HMD compact. 

The semiconductor laser used in our system was the SLM-RGB-T20-F manufactured by Sumitomo Electric, which emits red, green, and blue (RGB) laser light from a single package.
These light are sent from the SLM in sequence, and the Arduino receives the signals to illuminate a specific color, thus synchronizing the RGB signals from the SLM with the color of the laser beam. The SLM has a frame rate of 60 Hz and color-field-rate of 360 Hz, so the colors are switched six times per frame (Fig. \ref{fig:full_color}b).

\subsection{Theoretical FOV}
Regarding the FOV of our holo-HMD prototype described in Section \ref{sec:FOV}, the holo-HMD is designed with the focal length (25 mm) away from the concave mirror as the viewpoint. In this case, if a magnified reconstructed image is placed 250 mm from the viewpoint, the magnification ratio will be 10 times. The size of the enlarged reconstructed image that can be placed within the viewing area is limited to 80 mm, and the theoretical maximum FOV is $18.23^\circ$.

\subsection{Limit of phase-correction}
\begin{figure}[ht]
  \centering
    \includegraphics[clip,scale=0.6]{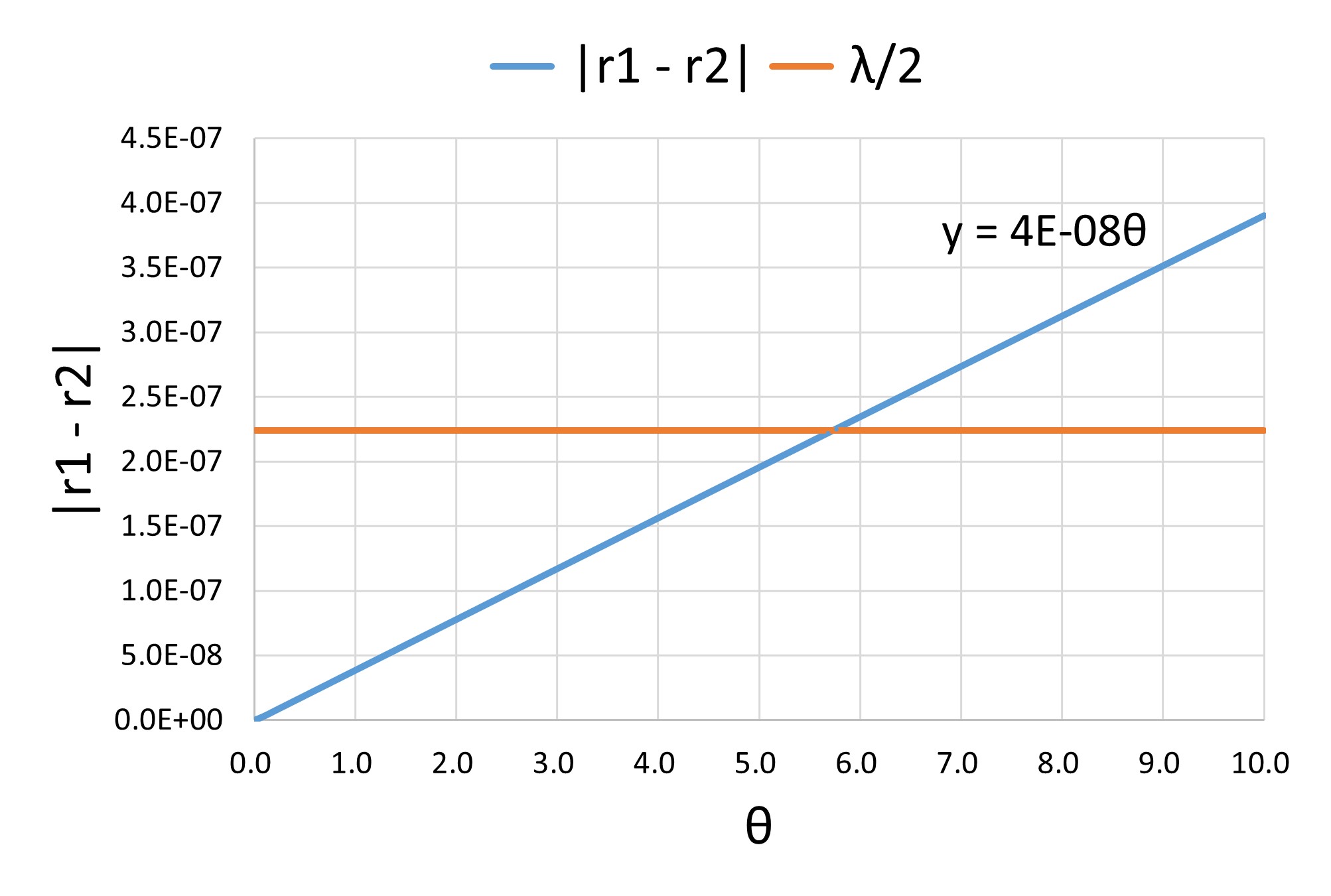}
    \caption{Maximum rotation angle $\theta$ satisfying sampling theorem.}
    \label{fig:path}
\end{figure}

For calculating object light phase-correction described in Section \ref{sec:phase-correction}, the relationship between the rotation angle $\theta$ and optical path difference $|r_1-r_2|$ is shown in Fig. \ref{fig:path} when the wavelength of light used for full-color reconstructing is blue ($\lambda=448$ nm). the $\theta$ satisfying the sampling theorem is determined as
\begin{equation}
   0^\circ < \theta < 5.77^\circ.
\end{equation}

\subsection{Assembly error}
When the reconstructed image was observed with our holo-HMD prototype, it was confirmed that the image was not obtained at the specified depth position. Since a concave mirror is used, the reconstructed image is formed once as a real image of the concave mirror, and the user observes the magnified virtual image. Therefore, if the position of the real image changes even slightly, the distance from the user's viewpoint to the virtual image changes, and the assembly error must be considered when checking the depth expression.

One of the causes of errors in the position of the real image is that the laser light is not perfectly collimated when it passes through the lens. Therefore, we measured the error of the real image position inside our holo-HMD prototype compared with the theoretical value.

\begin{figure}[ht]
  \centering
    \includegraphics[clip,scale=0.52]{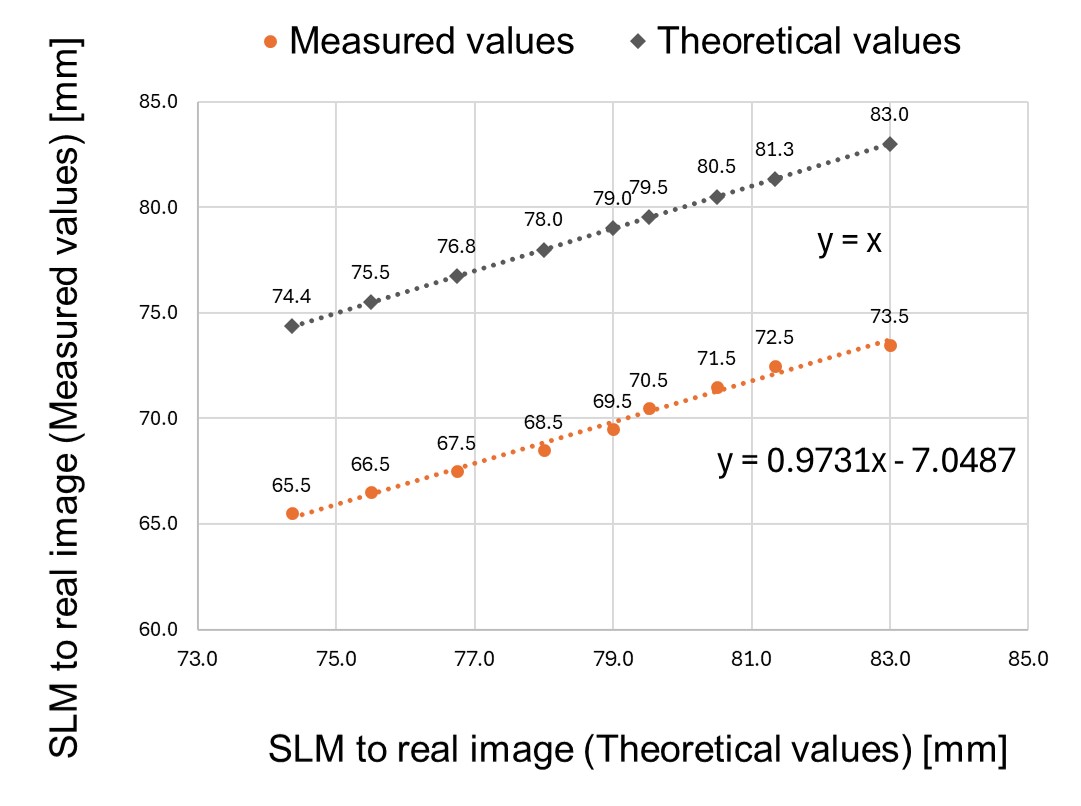}
    \caption{Position error of real image between theoretical and measured values of our holo-HMD prototype. The horizontal axis represents the distance from the SLM to the real image used in the calculation, while the vertical axis shows the measured distance at which the image was actually formed.}
    \label{fig:error}
\end{figure}

Fig. \ref{fig:error} shows the theoretical and measured values of the real image position in our holo-HMD prototype. The $x$-axis is the theoretical value of the real image position, and the $y$-axis is the measured value. With the optical system of our prototype, when the user wants to observe an virtual image magnified 10 x, for example, the theoretical position of the real image is 75.5 mm away from the SLM ($x$-axis). However, we tested our prototype, the image formation position for obtaining the same magnification rate was 66.5 mm away from the SLM ($y$-axis).

Fig. \ref{fig:error} also summarizes the theoretical and measured values of the actual image formation position for various magnification ratios, and the measured values are shown with the orange approximate straight line. The error between the theoretical and measured values was found to be about 9 mm.

\subsubsection{Coordinate transformation considering assembly errors}
As described in Section \ref{sec:concave mirror}, since our holo-HMD prototype uses a concave mirror, it is necessary to obtain the coordinates of the real image before magnification, which is necessary for CGH calculation, on the basis of the coordinates of the magnified image observed by the user. Therefore, we consider the coordinate transformation when considering the error in the position of the real image.

Assuming that the coordinates $P_i(x_i,y_i,z_i)$ of the image after magnification as observed by the user are the input parameters for CGH calculation, the theoretical coordinates $P_r(x_r,y_r,z_r)$ of the real image before magnification are the concave mirror coordinates in the transformation formula (Eq. \ref{eq:x_r,y_r,z_z}). Therefore, the final coordinates $P_f(x_f,y_f,z_f)$, taking into account the error from the theoretical value of the real image position, are as follows when $z_f$ is transformed using the approximate formula in Fig. \ref{fig:error}.
\begin{equation}\label{eq:x_f,y_f,z_f}
    P_f(x_f,y_f,z_f) = \left(x_r,y_r,0.9731z_r-7.0487\right)
\end{equation}

\section{Experiments}
\subsection{FOV}

\begin{figure}[ht]
    \begin{center}
        \begin{tabular}{cc} 
            \includegraphics[scale=0.2]{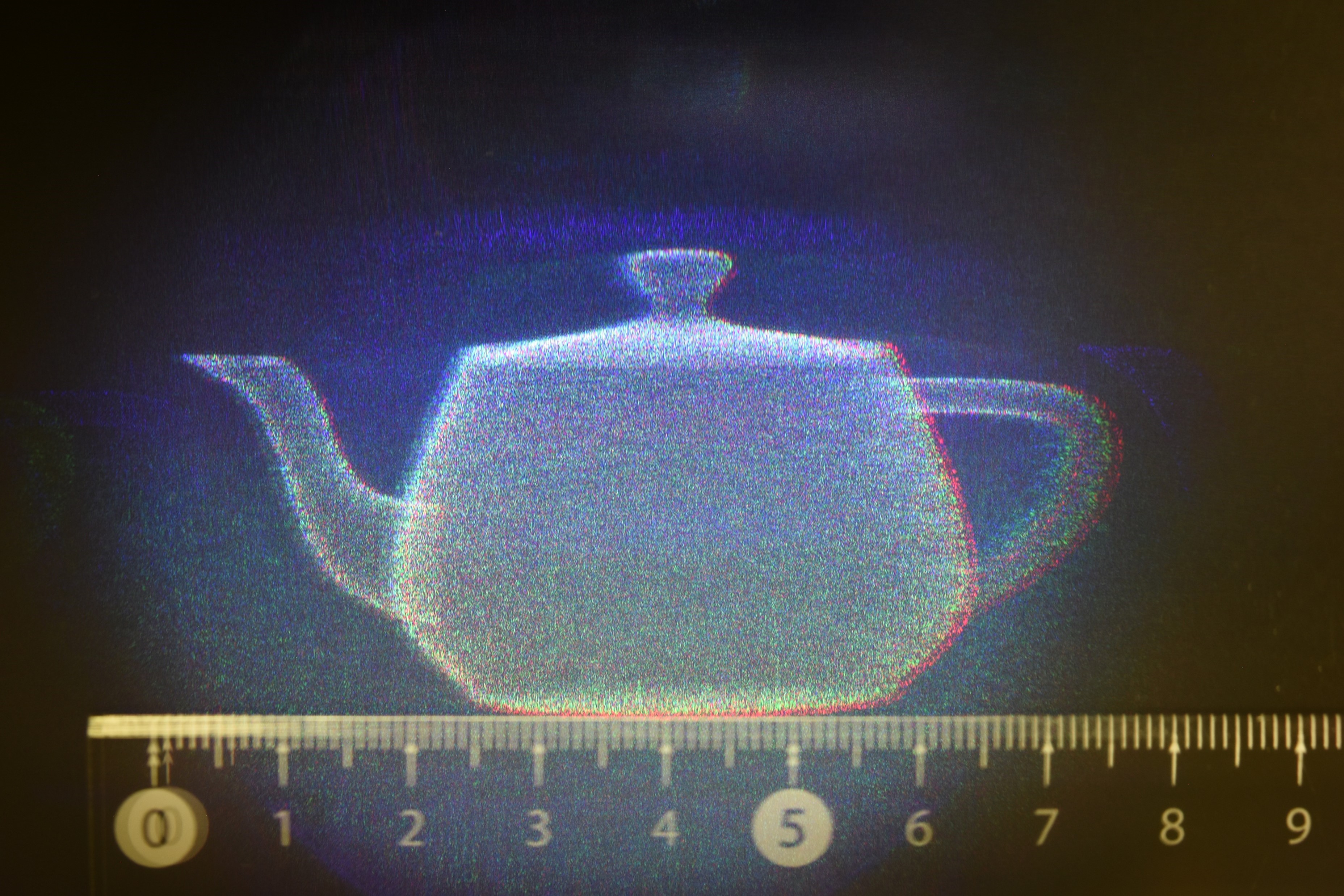} &
            \includegraphics[scale=0.2]{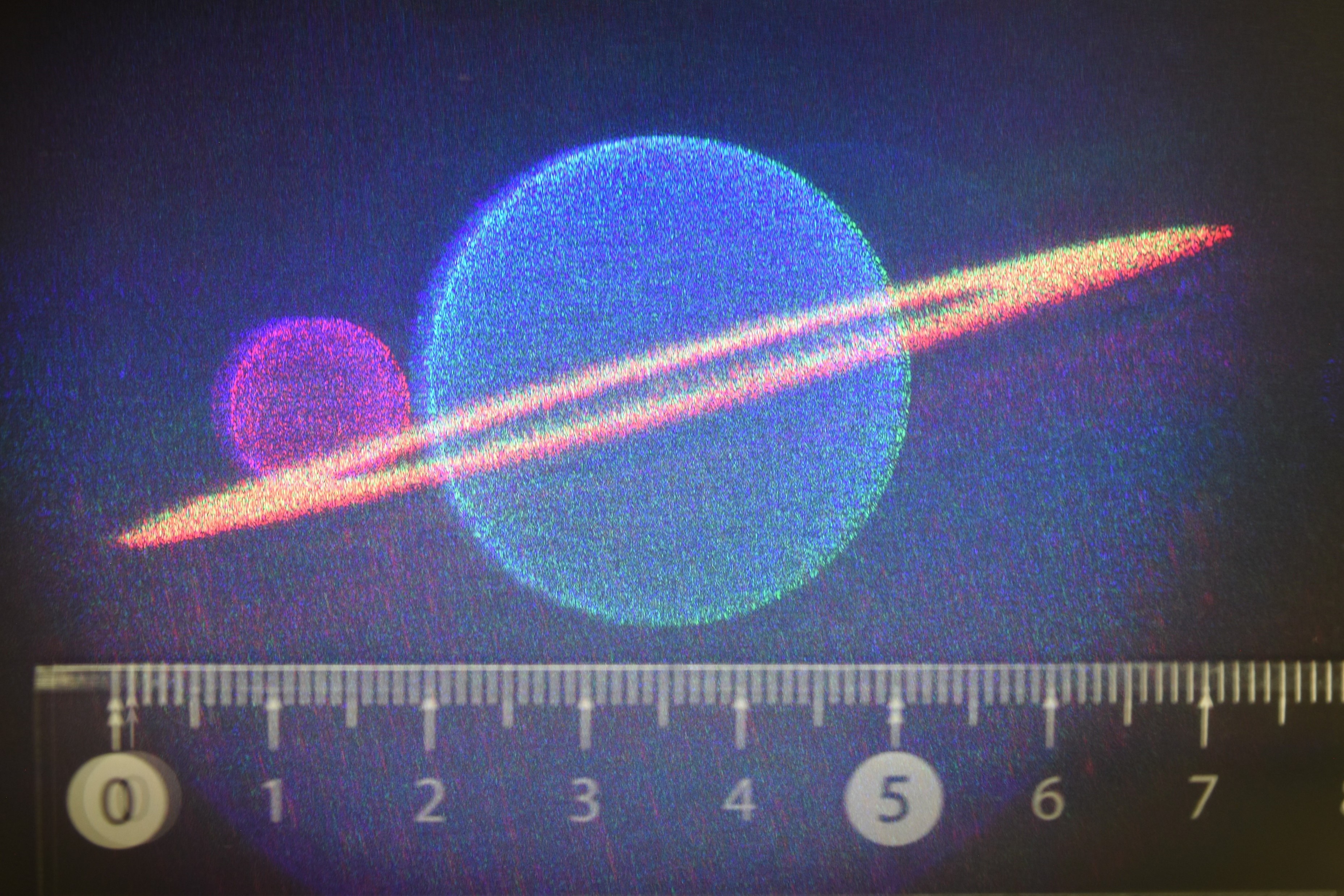} \\
            (a) & (b)
        \end{tabular}
    \end{center}
\caption{Reconstructed image: (a) FOV: size of the image is 75 mm from viewpoint, (b) full-color image: three different colors are displayed by combining two of R, G, and B.}
\label{fig:rec_image}
\end{figure}

A teapot was reconstructed at the maximum size that could be captured with the camera, 250 mm from the viewpoint. The reconstructed image is shown in Fig. \ref{fig:rec_image}a. The size of the captured image was 75 mm, and the calculated FOV angle was $17.06^\circ$. The theoretical value of the FOV angle described in Section \ref{sec:FOV} was $18.23^\circ$, resulting in an error of $-1.17^\circ$, or $-6.4\%$. Since our holo-HMD prototype reconstructs in full color, the FOV angle in the case of blue, which has the smallest FOV angle among the three primary colors of light, was targeted.

\subsection{Full-color image}
The results of the full-color reconstruction is shown in Fig. \ref{fig:rec_image}b. Three different colors are displayed by combining two of R, G, and B. The planet in the center is light blue (intermediate color between G and B), the satellite on the left is purple (intermediate color between R and B), and the ring of planets is yellow (intermediate color between R and G). 
No flicker due to the time division of the RGB lasers was observed because of the sufficient color refresh rate of 360 Hz in this full-color system.

\subsection{Depth expression}

The images were reconstructed at positions 50 and 250 mm in depth from the viewpoint. The polar bear and seal object were placed at each position, and the images were captured by focusing the camera on one object at a time. When the seal at 50 mm from the camera was focused on, the reconstructed image and polar bear at 250 mm from the camera were blurred. In contrast, when the polar bear 250 mm from the camera was focused on, the out-of-focus reconstructed image 50 mm away and the seal were blurred. This indicates that our developed holo-HMD prototype is capable of simultaneously expressing different depths.

\begin{figure}[ht]
    \begin{center}
        \begin{tabular}{cc} 
            \includegraphics[scale=1.1]{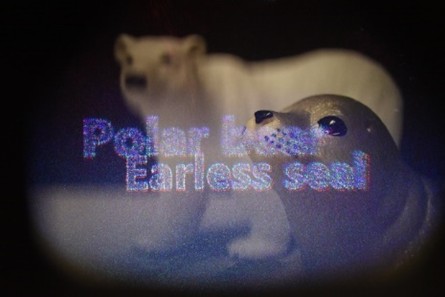} &
            \includegraphics[scale=1.1]{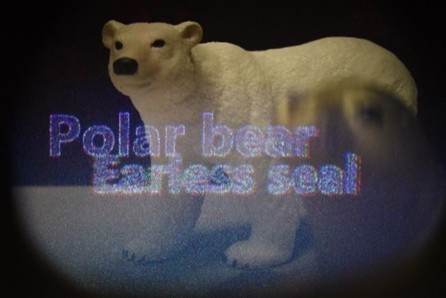} \\
            (a) & (b)
        \end{tabular}
    \end{center}
\caption{Depth expression: (a) focus on seal 50 mm from viewpoint, (b) focus on polar bear 250 mm from viewpoint.}
\label{fig:depth_rep}
\end{figure}

\subsection{six degrees of freedom}

\begin{figure}[ht]
    \begin{center}
        \begin{tabular}{ccc} 
            \includegraphics[scale=0.9]{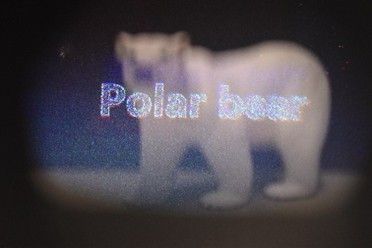} &
            \includegraphics[scale=0.9]{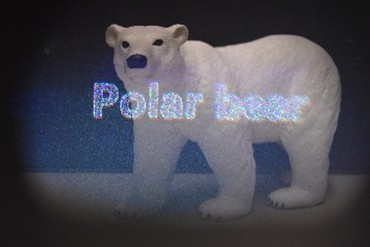} &
            \includegraphics[scale=0.9]{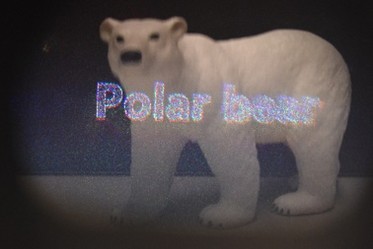} \\
            (a) & (b) & (c)
        \end{tabular}
    \end{center}
\caption{Reconstructed forward and backward shift images: (a) forward 150 mm, (b) reference position, (c) backward 150 mm.}
\label{fig:Forward and backward}
\end{figure}

\begin{figure}[ht]
    \begin{center}
        \begin{tabular}{ccc} 
            \includegraphics[scale=0.9]{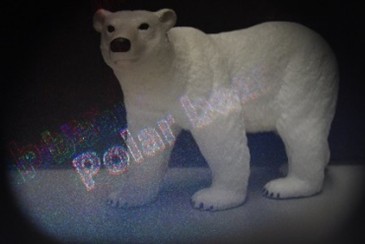} &
            \includegraphics[scale=0.9]{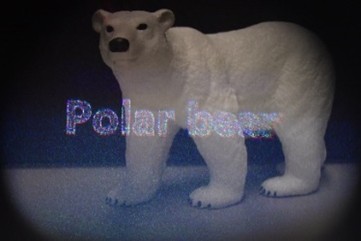} &
            \includegraphics[scale=0.9]{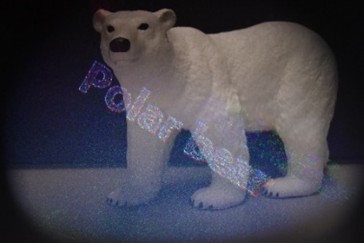} \\
            (a) & (b) & (c)
        \end{tabular}
    \end{center}
\caption{Reconstructed image of roll rotation: (a) $-30^\circ$, (b) $0^\circ$, (c) $30^\circ$.}
\label{fig:roll}
\end{figure}

\begin{figure}[ht]
    \begin{center}
        \begin{tabular}{ccc} 
            \includegraphics[scale=0.8]{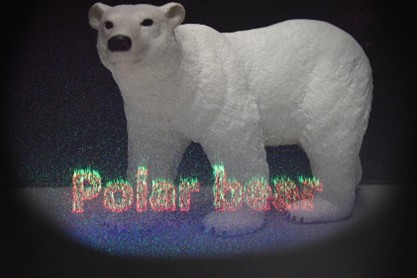} &
            \includegraphics[scale=0.8]{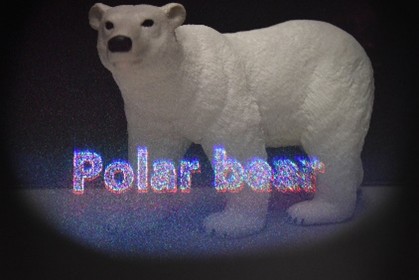} &
            \includegraphics[scale=0.8]{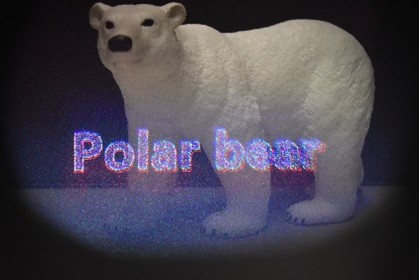} \\
            (a) & (b) & (c) \\
            \includegraphics[scale=0.8]{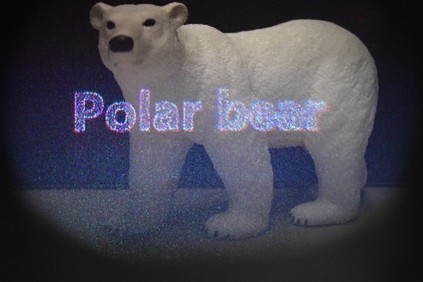} &
            \includegraphics[scale=0.8]{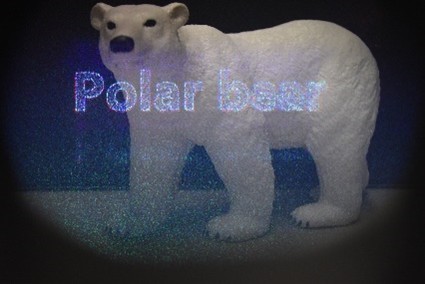} &
            \includegraphics[scale=0.8]{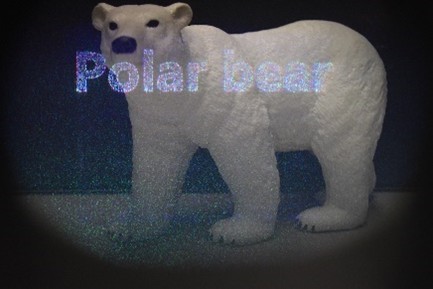} \\
            (d) & (e) & (f)
        \end{tabular}
    \end{center}
\caption{Reconstructed image of pitch rotation ($\fallingdotseq$ vertical shift): (a) $1.5^\circ$, (b) $1.0^\circ$, (c) $0.5^\circ$, (d) $-0.5^\circ$, (e) $-1.0^\circ$, (f) $-1.5^\circ$.}
\label{fig:pitch}
\end{figure}

\begin{figure}[ht]
    \begin{center}
        \begin{tabular}{ccc} 
            \includegraphics[scale=0.8]{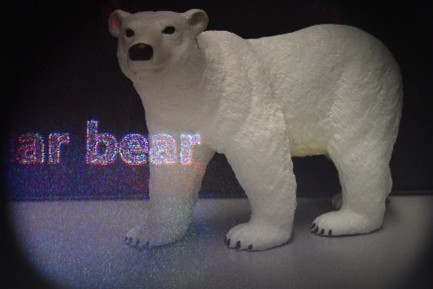} &
            \includegraphics[scale=0.8]{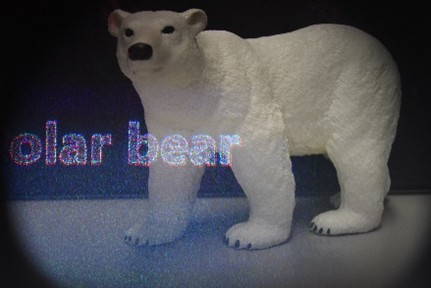} &
            \includegraphics[scale=0.8]{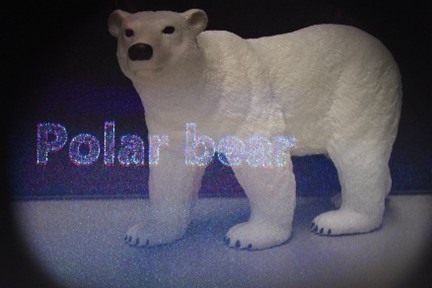} \\
            (a) & (b) & (c) \\
            \includegraphics[scale=0.8]{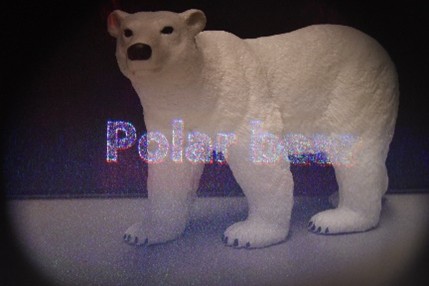} &
            \includegraphics[scale=0.8]{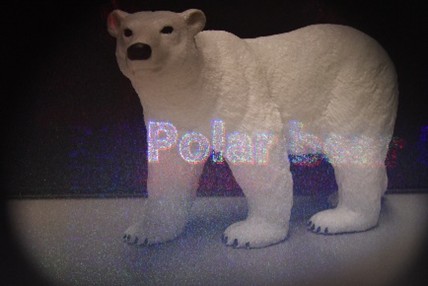} &
            \includegraphics[scale=0.8]{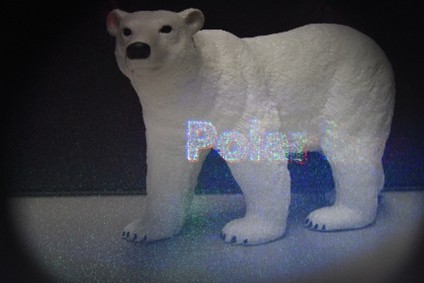} \\
            (d) & (e) & (f)
        \end{tabular}
    \end{center}
\caption{Reconstructed image of yaw rotation ($\fallingdotseq$ horizontal shift): (a) $-3^\circ$, (b) $-2^\circ$, (c) $-1^\circ$, (d) $1^\circ$, (e) $2^\circ$, (f) $3^\circ$.}
\label{fig:yaw}
\end{figure}

The six degrees of freedom reconstructed images with our holo-HMD prototype are shown in the following figures (Figs.\ref{fig:Forward and backward}-\ref{fig:yaw}). With our six degrees of freedom fast-calculation algorithm using phase-correction, the range of rotation and translation is narrow because the object light to be pre-calculated is flat, and the three degrees of freedom rotated and translated reconstructed images look the same. 

The reconstructed images were captured with a camera  fixed relative to the external environment at the viewpoint position of the holo-HMD, while the holo-HMD was rotated in roll rotation, moving back and forth, pitch rotation. 
As a result, the reconstructed images appear to rotate relative to the camera, indicating that the image is fixed with respect to the external environment.
\color{black}
Fig. \ref{fig:Forward and backward} shows reconstructed images when it was moved 150 mm forward and backward with reference to the polar bear, Fig. \ref{fig:roll} shows reconstructed images of roll rotation ($\pm30^\circ$), Fig. \ref{fig:pitch} shows reconstructed images of pitch rotation ($\pm0.5^\circ$, $1^\circ$, $1.5^\circ$) and Fig. \ref{fig:yaw} shows reconstructed images of yaw rotation ($\pm1^\circ$, $2^\circ$, $3^\circ$).
\color{black}

In the forward/backward shift images in Fig. \ref{fig:Forward and backward}, the reconstructed image moved forward and backward, and the polar bear was blurred when the focus was on the reconstructed image. The reconstructed image at each angle of the roll rotation in Fig. \ref{fig:roll} shows a tilted reconstructed image, but when the reconstructed image was rotated, a higher-order diffracted image was observed as shown in figures \ref{fig:roll} (a) and (b). This is thought to be the result of higher-order diffraction images that were cut off by the barrier at the viewpoint position, but were generated by the movement caused by the rotation. For the pitch rotation ($\fallingdotseq$ vertical shift) in Fig. \ref{fig:pitch} and yaw rotation ($\fallingdotseq$ horizontal shift) in Fig. \ref{fig:yaw}, the reconstructed image gradually shifted in each direction. The rotation angles for the pitch and yaw rotations were limited by the fact that the object light pre-calculated with the six degrees of freedom fast-calculation algorithm was flat and that the holograms were not aliased by resampling during the approximation process. Aliasing due to resampling must be prevented; therefore, it is difficult to expand the rotation angle.

\subsection{Calculation time}

\begin{figure}[ht]
  \centering
    \includegraphics[clip,scale=0.7]{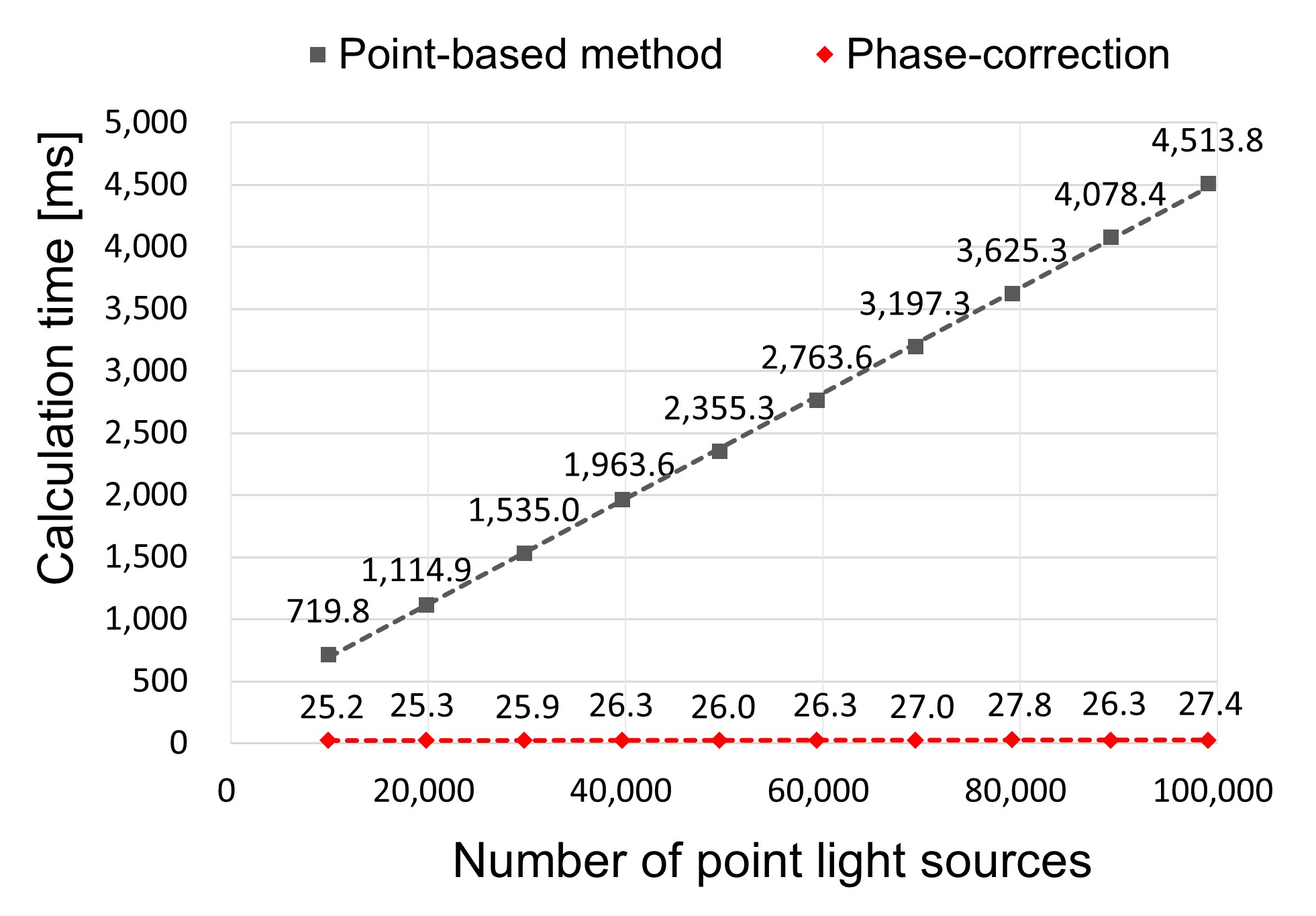}
    \caption{Calculation time comparison of phase-correction with point-light method.}
    \label{fig:calc-time}
\end{figure}

\begin{table}[ht]
    \centering
    \caption{Computer parameters}
    \vspace{1ex}
    \label{tab:Computer parameters}
    \begin{tabular}{lc}
        \hline
        CPU & Intel\textregistered{} Core\texttrademark{}i7-9700 (3.00[GHz])\\
        \hline
        GPU & NVIDIA GeForce RTX 2070 SUPER\\
        \hline
        Memory size & 16.0[GB] \\
        \hline
        OS & Windows 10 Pro 64bit \\
        \hline
        Pro. lang. & NVIDIA Cuda Ver.12.3 \\
        \hline
    \end{tabular}
\end{table}

Fig. \ref{fig:calc-time} compares the time from computation to display per CGH when using the phase-correction calculation with the point-light method for each number of point light sources $N$. These results are the average of the time taken to generate a CGH ten times. The parameters of the computer used to measure the computation time are listed in Table \ref{tab:Computer parameters}. The point-light method is computed in parallel using a GPU. The computational complexity of the point-light method is determined from the number of pixels in the hologram $P$ and $N$ as discussed in \ref{sec:algorithm}. The calculation results with the point-light method indicates that under a constant $P$, the computation time increases proportionally as $N$ increases. However, when using phase-correction, the computation time is almost constant even when $N$ increases, because the computation amount is determined only from $P$ not from $N$. The calculation results using phase-correction calculations were in the range of 25 to 28 ms, and the frame rate was about 35.7 to 40 fps. These results indicate that our holo-HMD prototype can calculate and display a CGH in real-time close to 40 fps, and that real-time display can be realized even for CGH calculation of more complex virtual objects.

\section{Discussion}
Real-time realization of rotational and translational motion with six degrees of freedom was successfully achieved, thereby addressing one of the most critical challenges in holo-HMDs.
In addition, the developed prototype incorporates all fundamental functions required for an HMD, including motion tracking, a wide FOV, and color display capability.
While most previous studies have focused on individual component technologies, our system integrates these functions into a single lightweight and compact package suitable for head-mounted use.
These results indicate that holo-HMDs have the potential to become practically viable in the near future.

However, several issues were also identified during the development of this system, which remain as subjects for future work.
The FOV of some AR glasses is in the range of $20^\circ$- $30^\circ$, and the prototype developed in this study achieves an FOV of approximately $20^\circ$, which is comparable to these devices. This indicates that our system has moved sufficiently close to the practical realization of holo-HMDs \cite{kishishita2014analysing}.
However, in recent years, AR glasses with wider FOVs of approximately over $50^\circ$ have become increasingly common. Furthermore, VR systems typically provide even wider FOVs over $90^\circ$. Therefore, it will be necessary to investigate optical systems capable of achieving a significantly wider FOV in future work.
The use of holographic optical elements (HOEs) for FOV expansion has been studied not only for increasing the FOV but also for enlarging the eyebox and enabling device miniaturization and weight reduction, making it a promising technology for future holographic displays.
In addition, further miniaturization and weight reduction can also be expected through the adoption of waveguide-based optical systems \cite{chen2025penta, gopakumar2024full, choi2025synthetic}.

In terms of image quality, significant degradation was observed due to speckle noise, further improvements are necessary \cite{yoneyama2023speckle}.
While DNN-based approaches require improvements in the characteristics of the reconstructed images, as discussed above in Section \ref{sec:fast}, they are highly promising in terms of their ability to reduce speckle noise.
In addition, color reproduction fidelity and objective image quality evaluation under speckle noise conditions remain unresolved issues, which will be addressed in future work.
\color{black}

\section{Conclusion}

We proposed a fast calculation algorithm for hologram-data generation for a holo-HMD supporting six degrees of freedom. The algorithm was implemented on a newly developed prototype HMD, demonstrating real-time hologram generation at 40 fps, with image updates responsive to head movements. 
The prototype with our optical system is small (125W $\times$ 130D$ \times$ 40H mm) and lightweight (287 g), making it suitable for head-worn operation.
The optical system has a $17.06^\circ$ FOV, ensuring a practical FOV for AR.
Optical experiments demonstrated that it is possible to display a full-color reconstructed image with the appearance of depth.

Overall, our holo-HMD prototype incorporates nearly all the essential features required for a practical holo-HMD system.
Furthermore, this work provides a useful guideline for the future development and practical implementation of holo-HMDs.

\section* {Disclosures}
There are no potential conflicts of interest, financial or otherwise, identified for this study.

\section* {Code and Data Availability}
The data that support the observations of this work would be made available by the corresponding
author upon reasonable request.

\section* {Acknowledgments}
These research results were obtained from the commissioned research (No. PJ012368C06801) by National Institute of Information and Communications Technology (NICT), Japan.


\bibliographystyle{unsrt}
\bibliography{report}   

\listoffigures
\listoftables

\end{document}